 \documentclass[twocolumn]{aa} 		
\usepackage{graphicx}
\def\gapprox{\lower.4ex\hbox{$\;\buildrel >\over{\scriptstyle\sim}\;$}}
\usepackage{txfonts}
\begin{document}

\title{		A Statistical Fractal-Diffusive Avalanche Model of a
		Slowly-Driven Self-Organized Criticality System }

\author{        Markus J. Aschwanden}

\institute{     Lockheed Martin Advanced Technology Center,
                Solar \& Astrophysics Laboratory,
                Org. ADBS, Bldg.252,
                3251 Hanover St.,
                Palo Alto, CA 94304, USA;
                e-mail: aschwanden@lmsal.com}

\date{Received 10 October 2011 / Accepted ... }

\abstract
{}
{We develop a statistical analytical model that predicts the occurrence 
frequency distributions and parameter correlations of avalanches in 
nonlinear dissipative systems in the state of a slowly-driven
self-organized criticality (SOC) system.} 
{This model, called the fractal-diffusive SOC model, is based on the
following four assumptions: (i) The avalanche size $L$ grows as a 
diffusive random walk with time $T$, following $L \propto T^{1/2}$;
(ii) The energy dissipation rate $f(t)$ occupies a fractal 
volume with dimension $D_S$, (iii) The mean fractal dimension
of avalanches in Euclidean space $S=1,2,3$ is $D_S \approx (1+S)/2$; and
(iv) The occurrence frequency distributions $N(x) \propto x^{-\alpha_x}$
based on spatially uniform probabilities in a SOC system are
given by $N(L) \propto L^{-S}$, with $S$ being the Eudlidean dimension. 
We perform cellular automaton simulations in three dimensions ($S=1,2,3$) 
to test the theoretical model.}
{The analytical model predicts the following statistical correlations: 
$F \propto L^{D_S} \propto T^{D_S/2}$ for the flux,
$P \propto L^{S} \propto T^{S/2}$ for the peak energy dissipation rate,
and $E \propto F T \propto T^{1+D_S/2}$ for the total dissipated energy; 
The model predicts powerlaw distributions for all parameters, with the 
slopes $\alpha_T=(1+S)/2$,  $\alpha_F=1+(S-1)/D_S$, $\alpha_P=2-1/S$, 
and $\alpha_E=1+(S-1)/(D_S+2)$.	The cellular automaton simulations
reproduce the predicted fractal dimensions, occurrence frequency 
distributions, and correlations within a satisfactory agreement within 
$\approx 10\%$ in all three dimensions.}
{One profound prediction of this universal SOC model is that the energy 
distribution has a powerlaw slope in the range of $\alpha_E=1.40-1.67$,
and the peak energy distribution has a slope of $\alpha_P=1.67$  
(for any fractal dimension $D_S=1,...,3$ in Euclidean space $S=3$),
and thus predicts that the bulk energy is 
always contained in the largest events, which rules out significant 
nanoflare heating in the case of solar flares.}

\maketitle

\section{       INTRODUCTION 			}

The statistics of nonlinear processes in the universe often shows
powerlaw-like distributions, most conspicously in energetic dynamic 
phenomena in astrophysics (e.g., solar and stellar flares, 
pulsar glitches, auroral substorms) and in catastrophic events in 
geophysics (e.g., earthquakes, landslides, or forest fires).
The most widely known example is the distribution of earthquake magnitudes,
which has a powerlaw slope of $\alpha \approx 2.0$ for the differential
frequency distribution (Turcotte 1999), the so-called Gutenberg-Richter (1954)
law.  Bak, Tang, and Wiesenfeld (1987, 1988) introduced the theoretical concept
of self-organized criticality (SOC), which has been initially applied to
sandpile avalanches at a critical angle of repose, and has been
generalized to nonlinear dissipative systems that are driven
in a critical state. Comprehensive reviews on this subject can be found
for applications in geophysics (Turcotte 1999), solar physics
(Charbonneau et al.~2001), and astrophysics (Aschwanden 2011).

Hallmarks of SOC systems are the scale-free powerlaw distributions of various
event parameters, such as the peak energy dissipation 
rate $P$, the total energy $E$, or the time duration $T$ of events.
While the powerlaw shape of the distribution
function can be explained by the statistics of nonlinear processes that have
an exponential growth phase and saturate after a random time interval
(e.g., Willis and Yule 1922; Fermi 1949; Rosner and Vaiana 1978;
Aschwanden et al.~1998; Aschwanden 2004, 2011), 
no general theoretical model has been developed that
predicts the numerical value of the powerlaw slope of SOC parameter
distributions. Simple analytical models that characterize the nonlinear
growth phase with an exponential growth time $\tau_G$ and the random
distribution of risetimes with an average value of $t_S$, predict a
powerlaw slope of $\alpha_P = 1 + t_S/\tau_G$ for the energy dissipation rate
(e.g., Rosner and Vaiana 1978; Aschwanden et al.~1998), but cellular
automaton simulations suggest a much more intermittent energy release
process than the idealized case of an avalanche with a single growth
and decay phase. An alternative theoretical
explanation for a slope $\alpha_E=3/2$ was put forward by a dimensional
argument (Litvinenko 1998), which can be derived from the definition of
the kinetic energy of convective flows, but this model entails a specific
physical mechanism that has not universal validity for SOC systems.

In this Paper we propose a more general concept where the powerlaw slope 
of the occurrence frequency distribution of SOC parameters depends on the 
fractal geometry of the energy dissipation domain.
We aim for a universal statistical model of nonlinear energy dissipation
processes that is independent of any particular physical mechanism.
The fractal structure of self-organized processes has been stressed 
prominently from beginning (Bak, Tang, \& Wiesenfeld 1987, 1988; 
Bak and Chen 1989), but no quantitative theory has been put forward
that links the fractal geometry to the size distribution of SOC events.
Fractals have been studied independently (e.g., Mandelbrot 1977, 1983, 1985),
while the fractal geometry of SOC avalanches was postulated 
(e.g., see textbooks of Bak 1996; Sornette 2004; Aschwanden 2011),
but no general self-consistent model has been attempted. 

This paper presents an analytical theory that derives
a theoretical framework to quantitatively link the concept of fractal
dimensions to the occurrence frequency distributions of SOC avalanche events
(Section 2), tests of the analytical theory with numerical simulations of
cellular automaton models in three Euclidean dimensions (Section 3),
a comparison and application to solar flares (Section 4), and a
summary of the model assumptions and conclusions (Section 5).

\section{	THEORY 			 		}

We derive in this Section a general model of the statistics of SOC processes,
but make use of a specific example of a SOC avalanche that is numerically 
simulated with a cellular automaton code and described in more detail in 
Section 3, to illustrate and validate our theoretical derivation. 

\subsection{	Diffusive Random Walk in Cellular Automaton Avalanches }			 
An avalanche in a cellular automaton model propagates via nearest-neighbor
interactions in random directions, wherever an unstable node is found.
The state of self-organized criticality ensures that the entire system
is close to the instability threshold, and thus every direction of
instability propagation is equally likely once a starting location
is triggered (if the re-distribution rule is defined to be isotropic).
We can therefore model the propagation of unstable nodes
with a random walk in a S-dimensional space, which has the characteristics
of a diffusion process and propagates in the statistical average a distance 
$x(t)$ that depends on the time $t$ as,
\begin{equation}
	x(t) \propto t^{1/2} \ .
\end{equation}

As a plausibility test we check this first assumption with an example of
a numerical simulation of a cellular automaton model that is described 
in more detail in Section 3. The simulated avalanche shown in Fig.~1 lasts
for a duration of 712 time steps and we show snapshots of the 
energy dissipation rate $de/dt$ in Fig.~1. 
The complete time evolution of the energy dissipation rate $de(t)/dt$, 
the total energy $e(t)$, the fractal dimension $D_2(t)$, and radius of 
the avalanche area $r(t)$ as a function of time is shown in Fig.~3. 
A movie of the avalanche that shows the time evolution for all 712 time steps
is included in the electronic supplementary data of this journal.
The movie illustrates also that the instantaneous propagation direction
of the avalanche is nearly isotropic, as we assumed here, regardless of
the prior evolution in the neighborhood of the instantaneous energy
release. Apparently, many of the grid points that have already been
touched by the avalanche previously are still in a meta-stable state
near the critical threshold that enables next-neighbor interactions.

We calculate the total time-integrated area $a(t)$ of the avalanche 
by summing up all unstable nodes where energy dissipation happened during
the time interval $[0,t]$ (counting each unstable node only once,
even when the same node was unstable more than once), and measure the 
mean radius $r(t)$ of the avalanche area by 
\begin{equation}
	r(t) = \sqrt{{a(t) \over \pi}} \ ,
\end{equation}
which closely follows the diffusive random walk distance $x(t) \propto
t^{1/2}$, as it can be seen in Fig.~3 (bottom right panel), or from the
circular area with radius $r(t)$ drawn around the starting point of
the avalanche in Fig.~1 (dashed circles around center marked with a cross).
Thus, the total time-integrated area $a(t)$ of an avalanche increases
with a diffusive scaling. If we define $T$ to be the total time duration 
of the avalanche, and $a_T=a(t=T)=\pi r^2(t=T)$ the final area of the 
avalanche, then the final linear size $L$ of the time-integrated avalanche is,
\begin{equation}
	L = \sqrt{a_T} = \sqrt{\pi} \ r(t=T) \ .
\end{equation}
According to the diffusive propagation we expect then not only a time 
evolution $r(t) \propto t^{1/2}$ (Eq.~1) for individual avalanches
(in the statistical average),
but also a statistical relationship between the size scales $L$ and the 
time durations $T$ for an ensemble of many avalanches, 
\begin{equation}
	L \propto T^{1/2} \ .
\end{equation}
The time-integrated area $a_T$ of an avalanche, as the outlined contours
in Fig.~1 show, appears to be a contiguous, space-filling area that is 
essentially non-fractal, although it has some ragged boundaries.
For an example of a 3-dimensional avalanche see McIntosh et al.~(2002),
which also shows essentially a space-filling 3-D topology for the
time-integrated avalanche volume.
This conclusion is somewhat intuitive in cellular automaton models, where 
avalanches can propagate over nearest neighbors only, and thus will
tend to cover a near space-filling area for isotropic propagation. 
Although the boundaries are somewhat ragged, the area $A$ is equivalent
to a circular area with a mean radius of $r(t)$, and thus the fractal
dimension would be approximately $D=2$, if a box-counting method 
is applied within a quadratic area with size $a_T=L^2=\pi r^2$. Thus, we define
that the total avalanche area has a Euclidean space-filling topology
within the diffusive boundary, and can be characterized
with a size scale $L$ and area $a_T=L^2$. Generalizing to 3-D space,
the volume $v$ of the avalanche boundary can be characterized by $v_T=L^3$.

\begin{figure*}
\centerline{\includegraphics[width=0.95\textwidth]{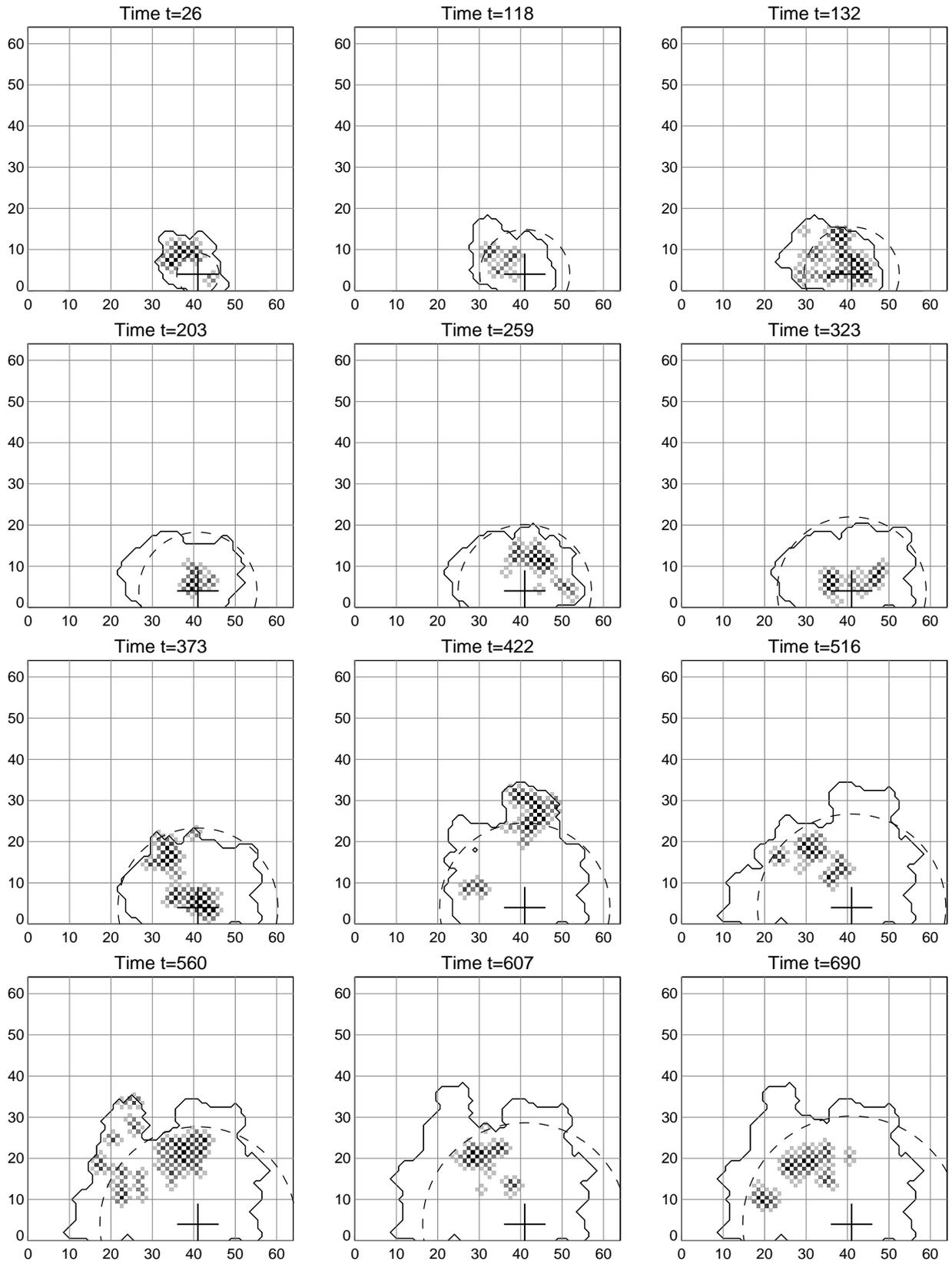}}
\caption{Time evolution of the largest avalanche event \#1628 in
the 2-D cellular automaton simulation with grid size $N=64^2$.
The 12 panels show snapshots at particular burst times from 
$t=26$ to $t=690$ when the energy dissipation rate peaked.
Active nodes where energy dissipation occurs at time $t$ are visualized
with black and grey points, depending on the energy dissipation level.
The starting point of the avalanche occurred at pixel $(x,y)=(41,4)$,
which is marked with a cross. The time-integrated envelop of the
avalanche is indicated with a solid contour, and the diffusive
avalanche radius $r(t) = t^{1/2}$ is indicated with a dashed circle.
The temporal evolution is shown in a movie available in the 
on-line version.}
\end{figure*}

\begin{figure}
\centerline{\includegraphics[width=0.5\textwidth]{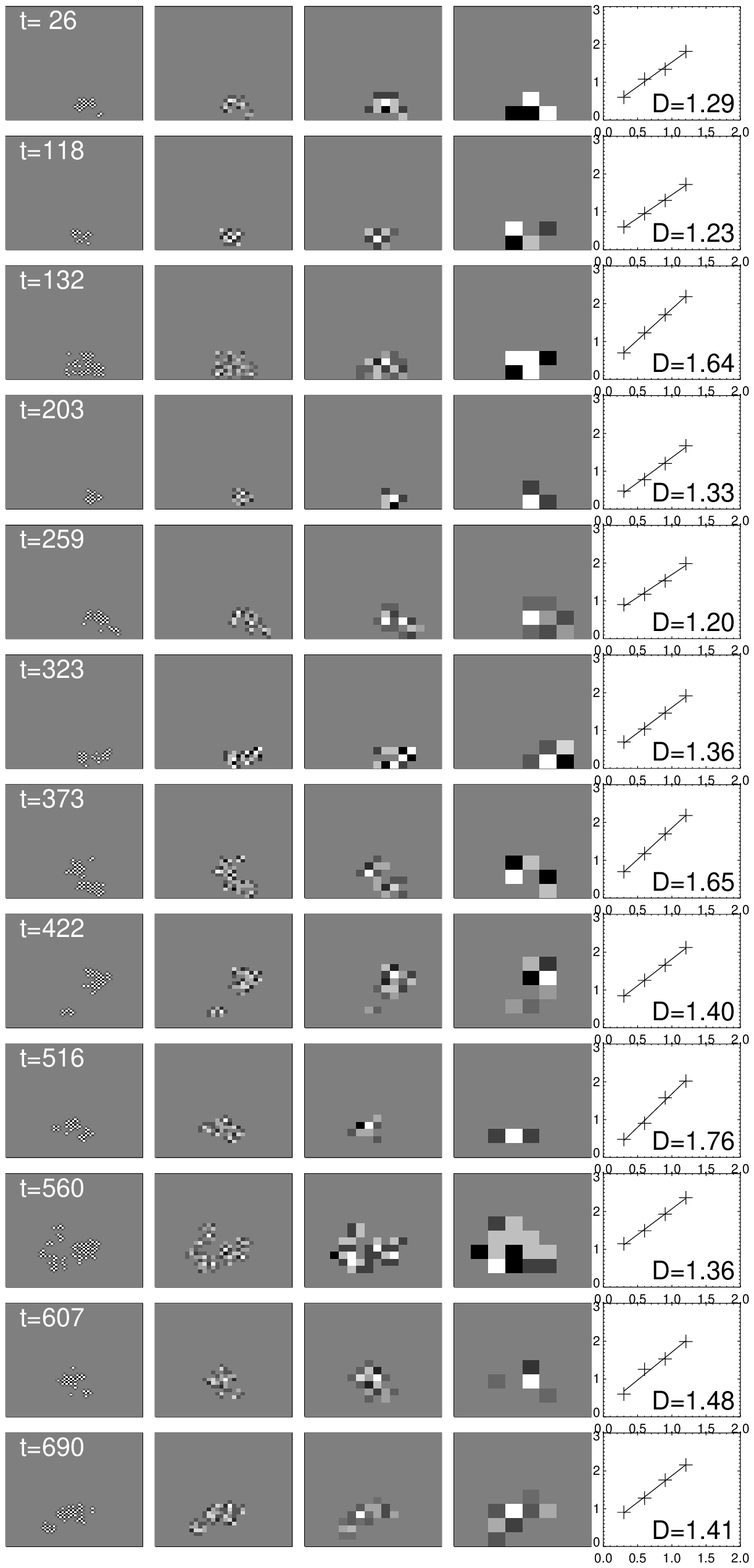}}
\caption{Determination of the fractal dimension $D_2=\log{A_i}/\log{x_i}$ 
for the instantaneous avalanche sizes of the 12 time steps of the
avalanche event shown in Fig.~1. Each row is a different time step
and each column represents a different binning of macropixels
($\Delta x_i=1,2,4,8$). The fractal dimension is determined by a linear
regression fit shown on the right-hand side. The mean fractal
dimension of the 12 avalanche snapshots is $D_2=1.43\pm0.17$.}
\end{figure}

\subsection{	The Fractal Geometry of Instantaneous Energy Dissipation  }

While we established the space-filling nature 
of the {\sl time-integrated} avalanche area 
with a (non-fractal) Euclidean dimension 
in the foregoing section, we will now, in contrast, derive the theorem that
the instantaneous area of energy dissipation in avalanches is fractal.
It is actually a key concept of SOC systems that the spatial structure of 
avalanches is {\sl fractal}. Bak and Chen (1989) express this most succintly 
in their abstract: {\sl ``Fractals in nature originate from self-organized 
critical dynamical processes''}. 

As it can be seen from the snapshots of an evolving avalanche shown in
Fig.~1, the instantaneous areas of energy dissipation cover a fraction
of the solid area $a(t)$ that is encompassed by the diffusive boundary.
A detailed inspection of the shapshots shown in Fig.~1 even reveals
a checkerboard pattern of instantaneous avalanche maps that emphasizes 
the fractal topology of cellular automaton avalanches. 
We make now the second major assumption that the area $A(t)$ of instantaneous
energy dissipation $de(t)/dt$ is fractal, or that the volume $V(t)$ 
for 3-D avalanches is fractal, respectively. To simplify the nomenclature,
we will generally refer to the fractal volume $V_S$ in S-dimensional
Euclidean space, which corresponds to $V_3=V$ for fractal volumes
in 3-D space, to $V_2=A$ for fractal areas in 2-D space, and
$V_1=X$ for fractal lengths in 1-D space. (Note that we use uppercase
symbols $V, A, X$ for fractal parameters, while we use lowercase symbols
$v, a, x$ for non-fractal Euclidean parameters).
A fractal volume $V_S$ can be defined by the {\sl Hausdorff dimension} 
$D_S$ in S-dimensional Euclidean space,
\begin{equation}
	D_S = \lim_{x\mapsto 0} {\log V_S \over \log x} \ ,
\end{equation}
where $V_S$ is the fractal volume with Euclidean scale $x$, or by the scaling
law,  
\begin{equation}
	V_S \propto x^{D_S} \ .
\end{equation}
In Fig.~2 we demonstrate the fractal nature of the 12 instantaneous avalanche
snapshots shown in Fig.~1. We rebin the avalanche area into macropixels
with sizes of $\Delta x_i=2^i, i=0,...,3$ (or $x=1,2,4,8$), measure the number
of macropixels $A_i$ that cover the instantaneous avalanche area, 
use the diffusive scaling $x(t) \propto t^{1/2}$ (Eq.~1) for the rebinned 
length units $x_i=x(t)/\Delta x_i$,
and determine the Hausdorff dimemsion $D_2$ from the linear regression fit
$\log(A_i)=D_2 \log(x_i)$. We find that that the 4 datapoints for each of
the 12 cases exhibit a linear relationship, which proves the fractality 
of the avalanche areas. The average fractal dimension of the 12 timesteps 
shown in Fig.~1 and 2 is $D_2 = 1.43 \pm 0.17$. 

\begin{figure*}
\centerline{\includegraphics[width=0.9\textwidth]{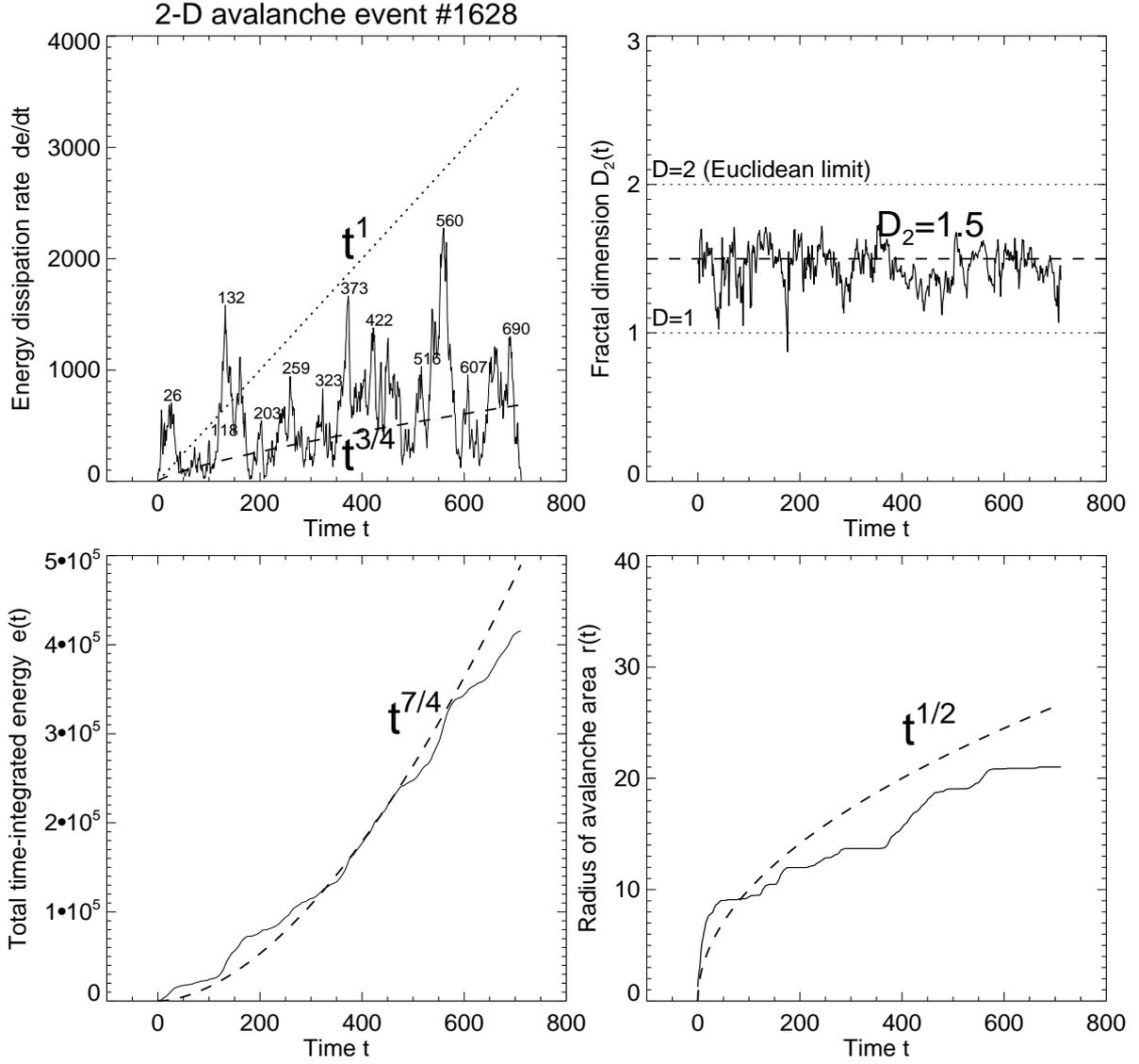}}
\caption{Time evolution of the largest avalanche event \#1628 in
the 2-D cellular automaton simulation with grid size $N=64^2$.
The time profiles include the instantaneous energy dissipation rate 
$f(t)=de/dt$ (top left),
the time-integrated total energy $e(t)$ (bottom left),
the instantaneous fractal dimension $D_2(t)$  (top right),
and the radius of the avalanche area $r(t)$ (bottom right).
The observed time profiles from the simulations are outlined in
solid linestyle and the theoretically predicted average evolution
in dashed linestyle. The statistically predicted values of the 
instantaneous energy dissipation rate $f(t) \propto t^{3/4}$ (dotted curve)
and peak energy dissipation rate $p(t) \propto t^1$ (dashed curve)
after a time interval $t$ are also shown (top left panel).
The 12 time labels from 26 to 690 (top left frame) correspond 
to the snapshot times shown in Fig.~1.}
\end{figure*}

We measure now the fractal dimension $D_2$ of the instantaneous energy 
dissipation volume in the 2-D avalanche for all 700 time steps of its
duration, shown for 12 time instants in Fig.~1. 
The unstable nodes (signifying instantaneous energy dissipation) 
are counted in each time step, which yield a number for the fractal volume 
or area $A(t)=V_2(t)$, while the size $x$ of the encompassing box is determined
from the area of the time-integrated avalanche, i.e., $x(t)=\sqrt{a(t)}$,
which yields the time evolution of the fractal dimension 
$D_2(t) = \log[V_2(t)]/\log[x(t)]$ (Eq.~5) as a function of time, 
shown in the top right panel in Fig.~3. The fractal dimension 
$D_2(t)$ fluctuates around a constant mean value of $D_2 = 1.45 \pm 0.13$, 
which is close to the arithmetic mean of the minimum dimension 
$D_{2,min} \approx 1$ and maximum Euclidean limit $D_{2,max}=2$, i.e. 
$\langle D_2 \rangle \approx (D_{2,min}+D_{2,max})/2=3/2$.
This corroborates our second major assumption that the instantaneous
volume of energy dissipation is fractal, and that the fractal dimension
can be approximated by a mean (time-independent and size-independent) 
constant during the evolution of avalanches, in the statistical average.

If we moreover define a mean energy dissipation rate quantum
$\langle \Delta E \rangle$ per unstable node, which is 
indeed almost a constant for a cellular automaton model
near the critical state, we expect a scaling of the instantaneous
dissipation rate (or flux) $f(t)$ that is proportional to the instantaneous 
dissipation volume $V_S$ (with Eq.~6),
\begin{equation}
	f(t) = {de(t) \over dt} \propto \langle \Delta E \rangle \ V_S(t) 
	                 = \langle \Delta E \rangle \ x(t)^{D_S} \ .
\end{equation}
Combining this with the diffusive expansion of the boundary
$x(t) \propto t^{1/2}$ (Eq.~1), we can then predict the average time 
evolution of the energy dissipation rate $f(t)=de(t)/dt$,
\begin{equation}
	f(t)= {de(t) \over dt} \propto 
	\langle \Delta E \rangle \ t^{(D_S/2)} \ .
\end{equation}

Integrating Eq.~(8) in time, we obtain the time evolution of the total
dissipated energy, $e(t)$, 
\begin{equation}
	e(t) = \int_0^t {de(\tau) \over d\tau} d\tau
		\propto \int_0^t \tau^{D_S/2} d\tau = t^{(1+D_S/2)} \ .
\end{equation}
Hence, for our 2-D avalanche (with $D_2=3/2$) we expect an evolution of 
$e(t) \propto t^{(7/4)}$, which indeed closely matches the actually 
simulated cellular automaton case, as we see in Fig.~3 (bottom left panel). 
The time evolution of the energy dissipation rate is shown in Fig.~3
(top left panel), which fluctuates strongly during the entire avalanche,
but follows in the statistical average the predicted evolution
$de(t)/dt \propto t^{D_2/2} = t^{(3/4)}$ for $D_2\approx 3/2$. 
Note that our analytical expressions of the time evolution of
avalanches, such as the linear size $x(t)$ (Eq.~1), the instantaneous
energy dissipation rate $f(t)$ (Eq.~7, 8), or the total dissipated
energy $e(t)$ (Eq.~9), do not predict the specific evolution of a
single avalanche event, but rather the statistical expectation value
of a large ensemble of avalanches, similar to the statistical nature of the 
diffusive random walk model (Eq.~1).

The time evolution of the instantaneous energy dissipation rate $f(t)$ 
fluctuates strongly, as it can be seen in for the largest avalanche 
simulated in a cellular automaton model (Fig.~3, top left panel).
We might estimate the peak values that can be obtained statistically
(after a time duration $t$) from the optimum conditions when the fractal
filling factor of the avalanche reaches a near-Euclidean filling,
i.e., in the limit of $D_S \mapsto S$. Replacing the fractal dimension
$D_S$ by the Euclidean limit $S$ in Eqs.~7 and 8 yields then (as an
upper limit) an expectation value for the peak $p(t)$, 
\begin{equation}
	p(t) \propto \langle \Delta E \rangle \ V_S^{max}(t) 
	     \propto \langle \Delta E \rangle \ x(t)^{S} 
	     \propto \langle \Delta E \rangle \ t^{(S/2)} \ .
\end{equation}

Denoting the energy dissipation rate in the statistical average after
time $t=T$ with with $F=f(t=T)$, the peak energy dissipation rate with
$P=p(t=T)$, and the total energy of the avalanche with $E=e(t=T)$, 
it follows from Eqs.~(8-10) that $E \propto F T = P^{D_S/S} T$, and
we expect then the following correlations between the three parameters 
$E$, $F$, $P$ and $T$ for an ensemble of avalanches,
\begin{equation}
	\begin{array}{l}
	 	E \propto T^{1+D_S/2}\\ 
	 	F \propto T^{D_S/2}  \\
	 	P \propto T^{S/2}    \\
	\end{array} \ . 
\end{equation}
For instance, for a 2-D avalanche with an average fractal dimension of
$D_2=3/2$ we expect the following two correlations,
$E \propto T^{7/4}$, $F \propto T^{3/4}$, and $P \propto T^1$ (see Fig.~3). 
The powerlaw indices for the correlated parameters are listed for the 
three Euclidean dimensions $S=1,2,3$ separately in Table 1. 

\begin{table*}
\caption{Theoretically predicted occurrence frequency
distribution powerlaw slopes $\alpha$ and power indices $\beta$
of parameter correlations predicted for SOC cellular automatons 
with Euclidean space dimensions $S=1,2,3$.}
$$
\begin{array}{p{0.5\linewidth}lllll}
\hline
\noalign{\smallskip}
Parameter	&Theory		&  S=1 	&  S=2 	&  S=3 		\\
\hline\noalign{\smallskip}
\hline
\hline\noalign{\smallskip}
Fractal Dimension: 			& D_S=(1+S)/2   	
&1	&3/2	&2\\ 
Length scale powerlaw slope: 		& \alpha_L=S  	
&1	&2	&3\\
Duration powerlaw slope: 		& \alpha_T=(1+S)/2  	
&1	&3/2	&2\\
Instantaneous energy dissipation rate slope: & \alpha_F=1+(S-1)/D_S  
&1      &5/3    &2      \\
Peak energy dissipation rate slope: 	& \alpha_P=2-1/S  
&1      &3/2    &5/3     \\
Energy powerlaw slope: 			& \alpha_E=1+(S-1)/(D_S+2)  
&1	&9/7	&3/2	\\ 
					&
&	&	&	\\
Diffusive scaling of length  L  with duration T, & L \propto T^{1/2}  	
& L \propto T^{1/2}  & L \propto T^{1/2}  &  L \propto T^{1/2}  \\
Correlation of peak rate  F  with duration  T , & F \propto T^{D_S/2}  	
& F \propto T^{1/2}  & F \propto T^{3/4}  &  F \propto T^{1}  \\
Correlation of peak rate  P  with duration  T , & P \propto T^{S/2}  	
& P \propto T^{1/2}  & P \propto T^{1}  &  P \propto T^{3/2}  \\
Correlation of energy  E  with duration  T , & E \propto T^{1+D_S/2}   
& E \propto T^{3/2}  & E \propto T^{7/4}  &  E \propto T^2  \\
\hline
\end{array}
$$
\end{table*}

\begin{figure*}
\centerline{\includegraphics[width=0.7\textwidth]{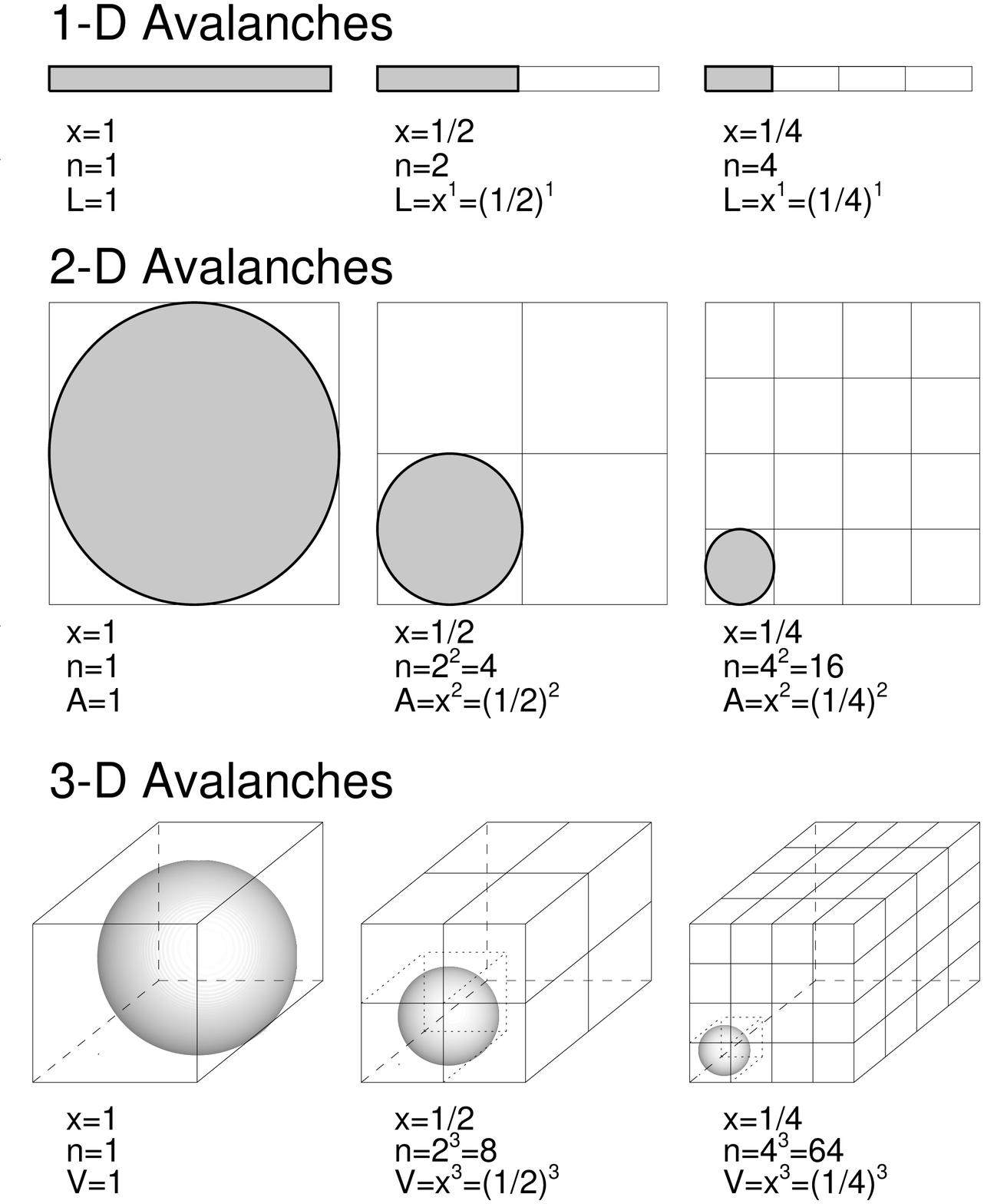}}
\caption{Schematic diagram of the Euclidean volume scaling 
of the diffusive avalanche boundaries, visualized as circles or spheres
in the three Euclidean space dimensions $S=1,2,3$. The Euclidean 
length scale $x$ of subcubes decreases by a factor 2 in each step 
($x_i=2^{-i}, i=0,1,2$), while the number of subcubes increases by 
$n_i=(2^i)^S$, defining a probability of $N(x_i) \propto x_i^{-S}$
for each avalanche size with size $x_i$.}
\end{figure*}

\subsection{	Occurrence Frequency Distributions 		}

Considering the probability of an avalanche with volume $V$, 
the statistical likelihood simply scales reciprocally to the volume
size $V$, if avalanches are equally likely in every space location
of a uniform volume $V_0$ of a system in a (self-organized) critical state. 
This is illustrated in Fig.~(4). For the 1-D Euclidean space,
$n=1$ avalanche can happen with the maximum size $L=L_0$ of the system
(top left), $n=2$ avalanches with the half size $L=L_0/2$, or
$n=4$ avalanches for a quarter size $L=L_0/4$. For the 2-D Euclidean
space, the number of possible avalanches that can be fit into 
the total area $A_0$ of the system is $n=1$ for $A=A_0$,
$n=2^2=4$ for $A=A_0/2$, or $n=2^4=16$ for $A=A_0/4$ (second row). 
For the 3-D Euclidean space we have, correspondingly, $n=1$ for $V=V_0$, 
$n=2^3=8$ for cubes of half size $L=L_0/2$, and $n=4^3=64$ for 
quarter-size cubes with $L=L_0/4$ (bottom row). 
So, generalizing to $S=1,2,3$ dimensions,
we can express the probability for an avalanche of size $L$ 
and volume $V_S=L^S$ as,
\begin{equation}
       N(L) \propto V_S^{-1} \propto L^{-S} \ .
\end{equation}
This simple probability argument is based on the assumption that the 
number or occurrence frequency of avalanches is equally likely throughout
the system, so it assumes a homogeneous distribution of critical states
across the entire system. 

The occurrence frequency distribution of length scales, 
$N(L) \propto L^{-S}$ (Eq.~12),
serves as a primary distribution function from which all other occurrence
frequency distribution functions $N(x)$ can the derived that have a
functional relationship to the primary parameter $L$. First we can
calculate the occurrence frequency distribution of avalanche time
scales, by using the diffusive boundary propagation relationship 
$L(T) \propto T^{1/2}$ (Eq.~4), by substituting the variable $T$ 
for $L$ in the distribution $N(L)$ (Eq.~12),
\begin{equation}
	N(T) dT = N(L[T]) \left| {dL \over dT} \right| dT 
	\propto T^{-[(1+S)/2]} \ dT \ .
\end{equation}
Subsequently we can derive the occurrence frequency distribution function
$N(F)$ for the statistically average energy dissipation rate $F=f(t=T)$ 
using the relationship $F(T) \propto T^{D_S/2}$ (Eq.~11),
\begin{equation}
	N(F) dF = N(T[F]) \left| {dT \over dF} \right| dF 
	\propto F^{-[1+(S-1)/D_S]} \ dF \ ,
\end{equation}
the occurrence frequency distribution function of the peak energy
dissipation rate $P$ using the relationship
relationship $P(T) \propto T^{S/2}$ (Eq.~11),
\begin{equation}
	N(P) dP = N(T[P]) \left| {dT \over dP} \right| dP 
	\propto P^{-[2-1/S]} 
	\ dP \ ,
\end{equation}
and the occurrence frequency distribution function
$N(E)$ for the total energy $E$ using the relationship
$E(T) \propto T^{1+D_S/2}$ (Eq.~11),
\begin{equation}
	N(E) dE = N(T[E]) \left| {dT \over dE} \right| dE 
	\propto E^{-[1+(S-1)/(D_S+2)]} 
	\ dE \ .
\end{equation}
Interestingly, this derivation yields naturally powerlaw functions for all
parameters $L$, $T$, $F$, $P$, and $E$, which are the hallmarks of SOC systems.
In summary, if we denote the occurrence frequency distributions 
$N(x)$ of a parameter $x$ with a powerlaw distribution with power index
$\alpha_x$,
\begin{equation}
	N(x) dx \propto x^{-\alpha_x} \ dx \ ,
\end{equation}
we have the following powerlaw coefficients $\alpha_x$ for the parameters
$x=T, F, P$, and $E$,
\begin{equation}
	\begin{array}{ll}
	\alpha_T &= (1+S)/2 \\ 
	\alpha_F &=  1+(S-1)/D_S \\ 
	\alpha_P &=  2-1/S \\ 
	\alpha_E &=  1+(S-1)/(D_S+2)\\	
	\end{array} \ .
\end{equation}
For instance, for our 2-D cellular automaton model with $S=2$ and
$D_S=3/2$ we predict powerlaw slopes of $\alpha_T=3/2=1.5$, 
$\alpha_F=5/3 \approx 1.67$, $\alpha_P=3/2=1.5$, 
and $\alpha_E=9/7 \approx 1.28$.
The powerlaw coefficients $\alpha_x$ are summarized in Table 1 
separately for each Euclidean dimension $S=1,2,3$. 

\subsection{Estimating the Fractal Dimension of Cellular Automatons}	

The dynamics of SOC systems is often simulated on computers with 
{\sl cellular automaton} codes, which have a S-dimensional lattice
of nodes, where a discretized mathematical redistribution rule is applied
once a local instability threshold is surpassed 
(e.g., Bak et al.~1987; Lu and Hamilton 1991; Charbonneau et al.~2001). 
Avalanches in such cellular automaton models propagate via nearest-neighbor 
interactions, which includes $(2 S+1)$ nodes in a S-dimensional lattice grid:
one element is the unstable node, and there are $2 S$ next neighbors.
For instance, a 1-dimensional grid in Euclidian space with dimension $S=1$ has
$(2S+1)=3$ nodes involved in a single nearest-neighbor relaxation step,
a 2-dimensional grid ($S=2$) has $(2S+1)=5$ nodes, and 
a 3-dimensional grid ($S=3$) has $(2S+1)=7$ nodes (Fig.~5).  

\begin{figure*}
\centerline{\includegraphics[width=0.8\textwidth]{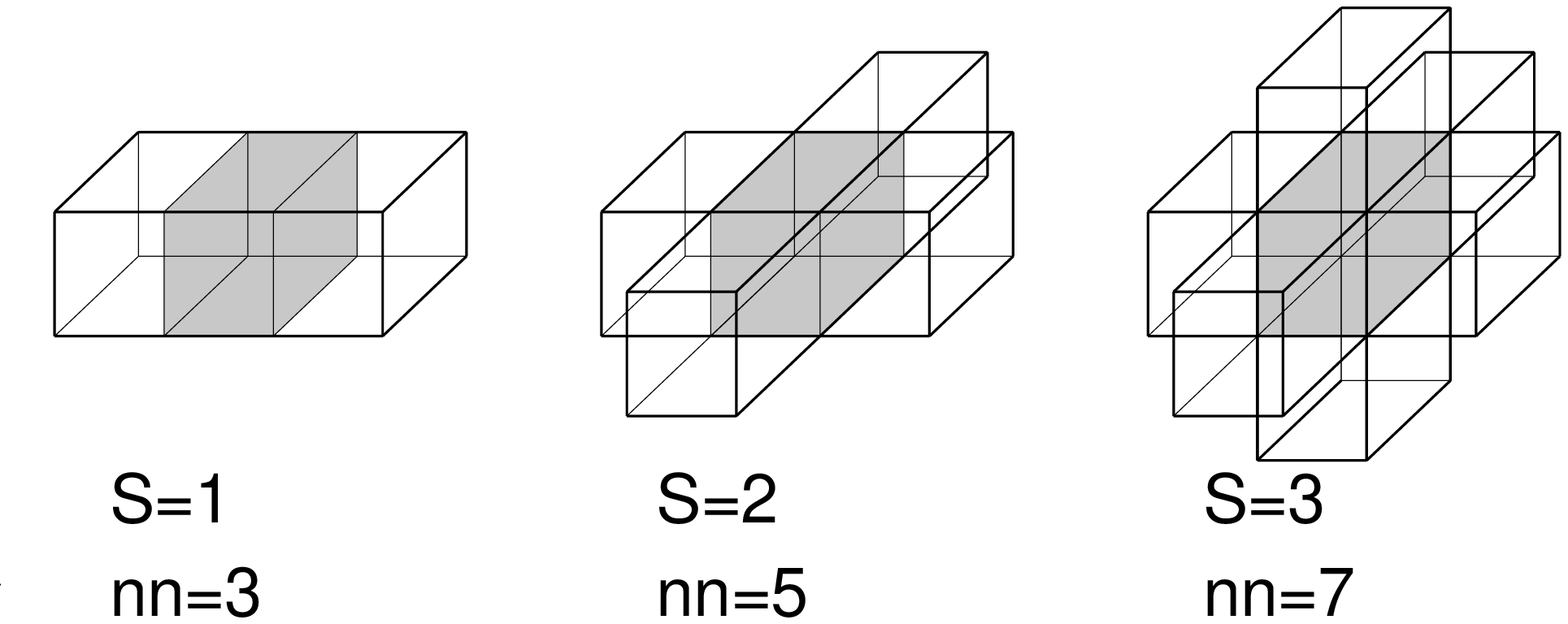}}
\caption{The fractal geometry of next-neighbor interactions are
shown for a cellular automaton lattice model for Euclidian
dimensions $S=1$ (left), $S=2$ (middle), and $S=3$ (right).
The unstable node is shaded with grey, and the next neighbor
nodes in white. The total number $nn$ of nodes involved in a
local redistribution rule scales as $nn=2 S + 1$ with the 
Euclidean dimension $S$.}
\end{figure*}

How can we estimate the fractal dimension for avalanches in such a 
lattice-based cellular automaton model? The minimum fractal dimension of
a growing 3-D avalanche corresponds to a 1-D linear structure, $D_{3,min}=1$, 
because avalanches evolve
by iterative propagation from one node to the next-neighbor node, so
a contiguous linear path from one node to the next neighbor is about the
sparsest spatial structure that still enables avalanche growth in a SOC model.
If the linear avalanche path would be discontinuous ($D_S < 1$),   
the avalanche is likely to die out due to a lack of unstable next neighbors,
so $D_{S,min} \approx 1$ represents practically a lower cutoff, which does
not exclude occasional values of $D_S < 1$ (fractal dust).
At the other extreme, when
all nodes are close to the instability threshold, an avalanche can
affect all next neighbors and grow as a nearly space-filling structure
with a maximum fractal dimension of $D_{S,max}=S$ that equals the
Euclidean space dimension $S$. 
The mean average fractal dimension $\langle D_S \rangle$ can then be
estimated from the geometric mean of the minimum $V_{S,min}$ and maximum 
fractal volume $V_{S,max}$, which is equivalent to the arithmetic mean 
of the minimum $D_{S,min}$ and maximum fractal dimension $D_{S,max}$
(Eq.~5),
\begin{equation}
	\langle D_S \rangle = {\log{ \langle V \rangle} \over \log{x}}  
	      = {\log{\sqrt{V_{S,min} V_{S,max}}} \over \log{x}} 
	      = {D_{S,min} + D_{S,max} \over 2} \ .
\end{equation}
From this we predict the following mean fractal dimensions $\langle D_S \rangle$
in different Euclidean spaces with dimensions $S=1,2,3$,
\begin{equation}
	\begin{array}{l}
		\langle D_1 \rangle = (1+1)/2 = 1   \\
		\langle D_2 \rangle = (1+2)/2 = 3/2 \\ 
		\langle D_3 \rangle = (1+3)/2 = 2   \ . 
	\end{array}
\end{equation}
or more generally as a function of the Euclidean space dimension $S$,
\begin{equation}
		\langle D_S \rangle \approx {(1+S) \over 2} \quad {\rm for}\ S=1,2,3 \ .
\end{equation}
Using these estimates of the mean fractal dimension $\langle D_S \rangle$
into Eq.~(18) we obtain the numerical values given in Table 1.
Note, that this estimate of the mean fractal dimension is only an
approximation, because the lower limit $D_{S,min} \approx 1$ is not
rigorously derived from probability theory. However, as the example
in Fig.~3 shows, the estimated fractal dimension of $D_2 = 1.5$
comes close to the observed mean value of 
$\langle D_2 \rangle = 1.45 \pm 0.13$ in this avalanche.

SOC systems that are different from the isotropic cellular automaton 
model we are using here may have different fractal dimensions. Thus our
theoretical prediction of the mean fractal dimension applies only to 
SOC systems with similar
isotropic next-neighbor redistribution rules. For any observed SOC system,
the fractal dimension $D_S$ can be empirically determined by measuring the
powerlaw slopes $\alpha_F$ or $\alpha_E$, which are a function of the
fractal dimension $D_S$ (Eq.~18).

\section{CELLULAR AUTOMATON SIMULATIONS} 

\subsection{Numerical Cellular Automaton Code}

An isotropic cellular automaton model that mimics a SOC system was 
originally conceived by Bak et al.~(1987, 1988) and first applied to
solar flares by Lu and Hamilton (1991). A version generalized to
$S=1,2,3$ dimensions is given in Charbonneau et al.~(2001) and is also
summarized in Aschwanden (2011). 

The numerical simulations of cellular automaton models take place in 
a S-dimensional cartesian grid, where nodes are localized by 
discretized coordinates, say with $x_{ijk}$ in $S=3$, and a physical
scalar quantity $B_{ijk}=B(x=x_{ijk})$ is assigned to each node. 
The dynamics of the system is initiated by quantized energy inputs
$\delta B(t)$ at random locations $x_{ijk}$ with a constant rate as
a function of time. At each time step $t$ and spatial node $x_{ijk}$,
the local stability is checked by measuring the local S-dimensional
{\sl ``curvature''} (Charbonneau et al.~2001) with
respect to the next neighbor cells (nodes) $x_{nn}=x_{i\pm1,j\pm1,k\pm1}$,
\begin{equation}
       \Delta B_{ijk} = B_{ijk} - {1 \over 2S} \sum B_{nn} \ .
\end{equation}
Most nodes are stable at a particular time if the system is driven slowly,
say with an input rate $\delta B/\langle B_{ijk}\rangle \ll 1$
(Charbonneau et al.~2001). The system is defined
to be stable, as long as the local gradients are smaller than some
critical threshold value $B_c$. However, once a local gradient
exceeds the critical value, i.e., $\Delta B_{ijk} \ge B_c$, a mathematical
redistribution rule is applied that smoothes out the local gradient
and makes it stable again. The redistribution rule simply spreads the
difference $\Delta B_{ijk}$ (or the threshold $B_c$) equally to the next 
neighbors (in an isotropic cellular automaton model),
\begin{equation}
       {\bf B}_{ijk} \mapsto {\bf B}_{ijk} - {2 S \over 2 S + 1} {\bf B_c} \ ,
       \quad
       {\bf B}_{nn} \mapsto {\bf B}_{nn} + {1 \over 2 S + 1} {\bf B_c} \ .
\end{equation}
Note that the amount of the redistributed quantity is the threshold energy 
$B_c$ in the models of Lu et al.~(1993) and
Charbonneau et al.~(2001), while it is the (larger) amount of the actual 
gradient $\Delta B_{ijk}$ in the original model of Lu \& Hamilton (1991).
If the node is unstable, then the actual gradient is larger than the
critical gradient, rather than smaller. This was modified in a later paper 
(Lu et al.~1993) by redistributing the threshold gradient rather than the 
full gradient, presumably due to numerical instabilities (Liu et al.~2002).
This redistribution rule is {\sl conservative}, in the sense that the 
quantity $B$ is conserved after every redistribution step, because the same 
amount is transferred to the next neighbors that is taken away from the 
central cell. However, although the scalar field quantity $B$ is conserved, 
the energy $B^2$ is not conserved after a redistribution step, because of 
the nonlinear (quadratic) dependence assumed, which was introduced in 
analogy to the magnetic field energy density $E_{mag}=B^2/8\pi$.
In fact, every redistribution of 
$|\Delta B| > B_c$ dissipates energy from the system, by an amount of
(Charbonneau et al.~2001),
\begin{equation}
	E = {2 S \over 2 S + 1} \left( {2 |\Delta B| \over B_c} - 1 \right) 
		B_c^2 \ .
\end{equation}
Thus, the minimum amount of dissipated energy is for $|\Delta B| \gapprox B_c$,
when the threshold $B_c$ is infinitesimally exceeded by $|\Delta B|$,
\begin{equation}
	E_{min} = {2 S \over 2 S + 1} B_c^2 \ .
\end{equation}
Once a node $x_{ijk}$ is found to be unstable, the check of unstable cells 
propagates to the next neighbors $x_{nn}=x_{i\pm1,j\pm1,k\pm1}$ in the next 
time step and all unstable neighbor cells are subject to the redistribution 
rule, and progressively continues to the next neighbors each time step
until all cells are stable again. Such a chain reaction of next-neighbor 
redistributions is called an {\sl avalanche event} (see examples in Fig.~4).  
Note that a minimum avalanche has to include at least one redistribution step.

In this study the author coded independently such a cellular automaton 
algorithm according to the specifications given in Section 2.5 of
Charbonneau et al.~(2001) and it was run with exactly the same system
parameters, such as the input quantity $\sigma_1 \le \delta B \le
\sigma_2$ homogeneously distributed in the range of 
$\sigma_1=-0.2$ to $\sigma_2=0.8$, the threshold quantity
$B_c=5$, with grid sizes of $N=128$ and 256 in one dimension ($S=1$),
$N=32$ and 64 in two dimensions ($S=2$), and $N=16$ and 24 in three dimensions
$(S=3)$. We sampled the total volumes $V$, energies $E$, peak energy
dissipation rate $P$, and durations $T$ of avalanches and were able to 
reproduce the results given in Charbonneau et al.~(2001) consistently, although
we used different random generators for the input, different time intervals
for the initiation phase and onset of SOC
($t_{SOC}= 3\times 10^6$ for $N=128$ and $S=1$;
$t_{SOC}=15\times 10^6$ for $N=256$ and $S=1$;
$t_{SOC}= 5\times 10^6$ for $N= 32$ and $S=2$;
$t_{SOC}=41\times 10^6$ for $N= 64$ and $S=2$;
$t_{SOC}= 4\times 10^6$ for $N= 16$ and $S=3$;
$t_{SOC}=15\times 10^6$ for $N= 24$ and $S=3$),
and slightly different powerlaw fitting procedures. In addition, we
calculated also the fractal dimensions of the SOC avalanches from the
avalanche volumes $V$ and the size $x$ of the enveloping Euclidean cube $x^S$,
\begin{equation}
	D_S = {\log{(V)} \over \log{(x)}} \ ,
\end{equation}
where $x$ is the largest spatial scale that brackets the fractal avalanche
volume in each spatial direction $(x,y,z)$ of the S-dimensional Euclidean
space. This definition is slightly different from the ``radius of gyration''
method employed in Charbonneau et al.~(2001), but follows more the standard
convention of fractal dimensions measured with box-counting methods.
In the following we show the results from two runs in each dimension 
$S=1,2,3$ and compare them with the theoretical predictions made in 
Section 2. 

\begin{figure*}
\centerline{\includegraphics[width=0.9\textwidth]{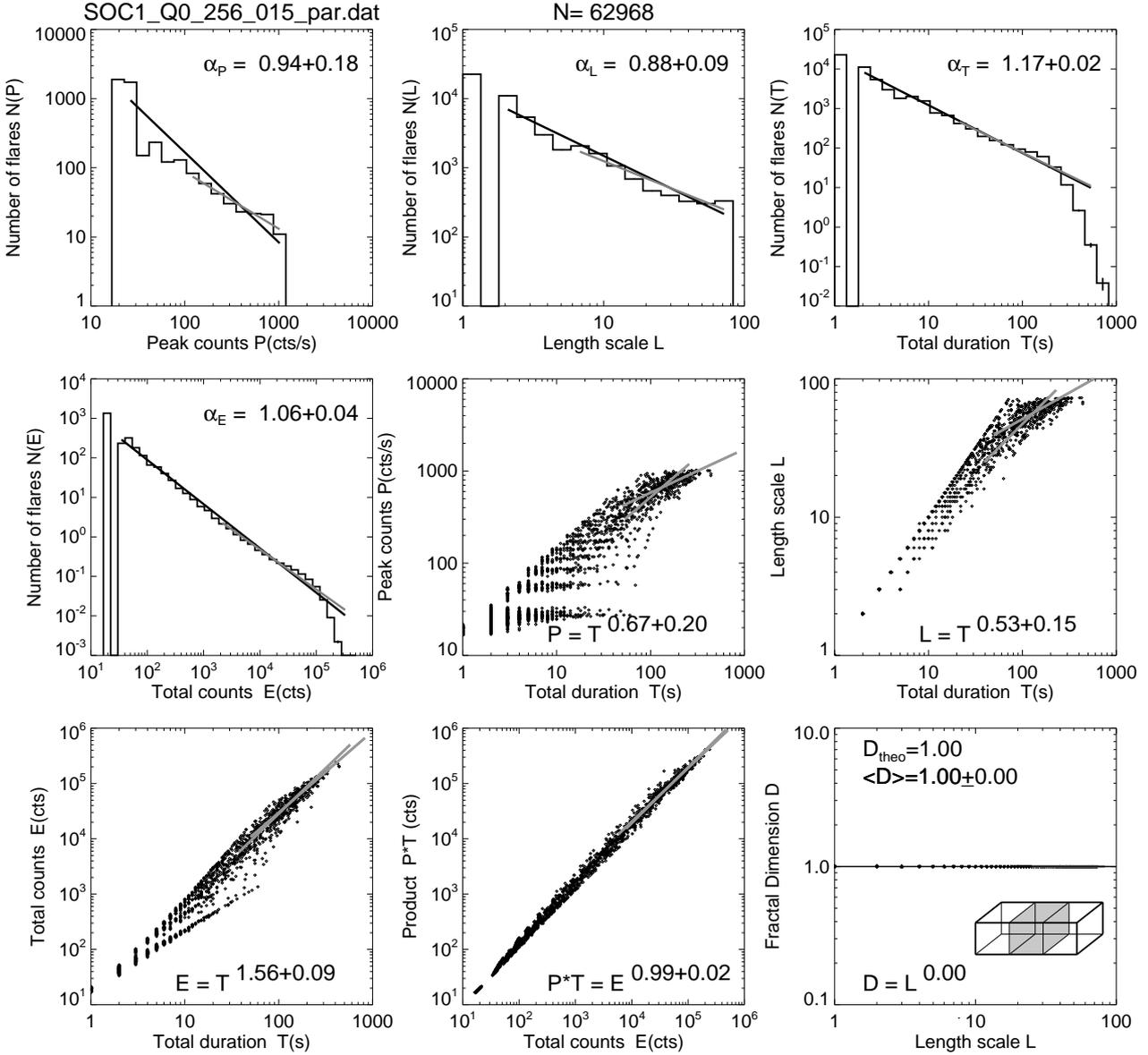}}
\caption{Cellular automaton simulations with a $N=256$ 1-D lattice
produced by a numerical code according to Charbonneau et al.~(2001).
The frequency distributions of peak energy dissipation rate $P$, total
energies $E$, time durations $T$, and fractal avalanche volumes $V$
are shown along with the fitted powerlaw slopes. Correlations between
the fractal dimension $D_1$ and parameters $E$, $P$, and $T$ are also
shown, fitted in the ranges of $P \ge 50$, $E \ge 50$, $T \ge 5$, 
and $V\ge 2$. Only a representative subset of 2000 events are plotted
in the scatterplots.}
\end{figure*}

\begin{figure*}
\centerline{\includegraphics[width=0.9\textwidth]{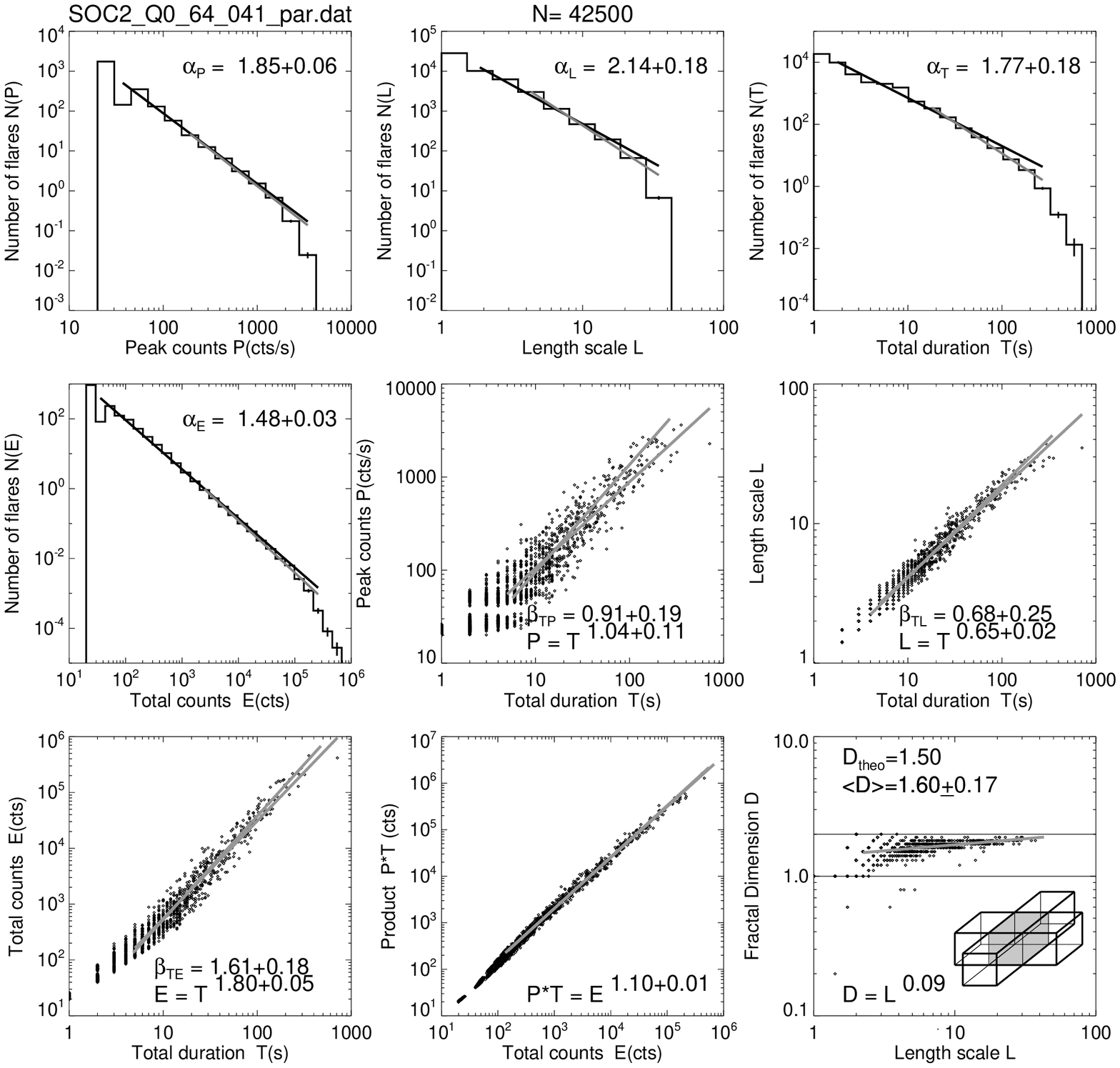}}
\caption{Cellular automaton simulations with a $N=64^2$ 2-D lattice
produced by a numerical code according to Charbonneau et al.~(2001).
Representation otherwise similar to Fig.~6.}
\end{figure*}

\begin{figure*}
\centerline{\includegraphics[width=0.9\textwidth]{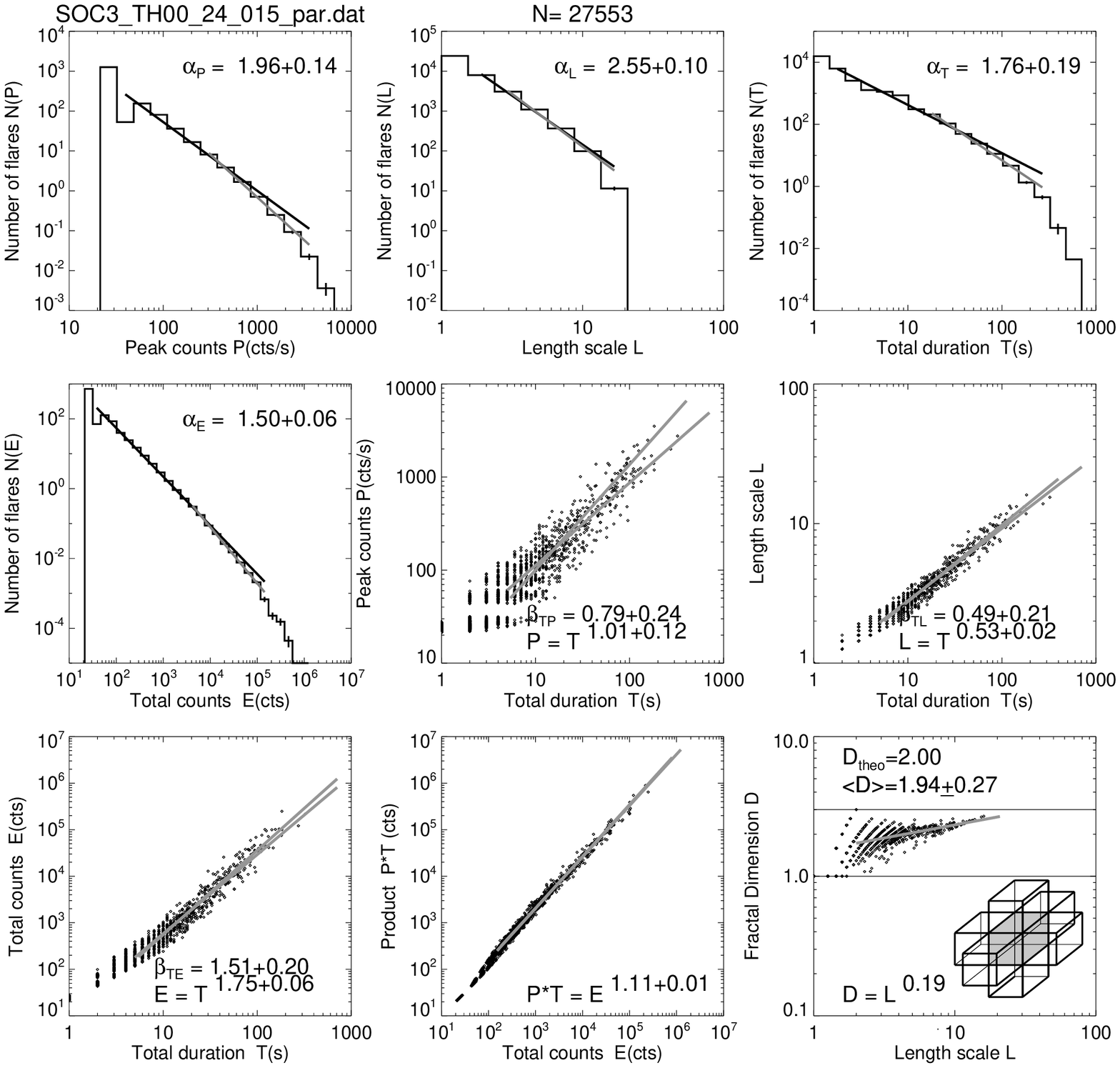}}
\caption{Cellular automaton simulations with a $N=24^2$ 3-D lattice
produced by a numerical code according to Charbonneau et al.~(2001).
Representation otherwise similar to Figs.~6 and 7.}
\end{figure*}

\begin{table*}	
\caption{Theoretically predicted and numerically simulated 
powerlaw slopes of occurrence frequency distributions of cellular
automaton models with Euclidean dimension $S=1,2,3$ in the state
of self-organized criticality.}
$$
\begin{array}{p{0.5\linewidth}lllll}
\hline\noalign{\smallskip}
Reference	& Theory  & S=1 	& S=2 	& S=3 	\\
\hline\noalign{\smallskip}
\hline
\hline\noalign{\smallskip}
\underbar{Fractal Dimension: $D_S$ }&	&		&		\\
Theory: $D_S$=(1+S)/2 	&{\bf 1.00}	&{\bf 1.50}	&{\bf 2.00}	\\ 
Simulation (N=256,64,24)&1.00 \pm 0.00	&1.60 \pm 0.17	&1.94 \pm 0.27	\\
Simulation (N=128,32,16)&1.00 \pm 0.00	&1.62 \pm 0.18	&1.97 \pm 0.29	\\
Charbonneau et al.~(2001)&		&1.58 \pm 0.02	&1.78 \pm 0.01	\\
McIntosh et al.~(2002)	&		&1.58 \pm 0.03	&1.78 \pm 0.01	\\
\underbar{Length scale powerlaw slope:  $\alpha_L$ }&&	&		\\
Theory: $\alpha_L$=S     &{\bf 1.00}	&{\bf 2.00}	&{\bf 3.00}	\\
Simulation (N=256,64,24)&0.88 \pm 0.09	&2.14 \pm 0.18	&2.55 \pm 0.10	\\
Simulation (N=128,32,16)&0.89 \pm 0.10	&1.96 \pm 0.09	&2.58 \pm 0.11	\\
\underbar{Energy powerlaw slope:  $\alpha_E$ }&&	&		\\
Theory: $\alpha_E$       &{\bf 1.00}	&{\bf 1.28}	&{\bf 1.50}	\\
Simulation (N=256,64,24)&1.06l\pm 0.04	&1.48 \pm 0.03	&1.50 \pm 0.06	\\
Simulation (N=128,32,16)&1.09 \pm 0.05	&1.42 \pm 0.03	&1.51 \pm 0.06	\\
Charbonneau et al.~(2001)&		&1.42 \pm 0.01	&1.47 \pm 0.02	\\
McIntosh et al.~(2002)	&		&1.41 \pm 0.01	&1.46 \pm 0.01	\\
\underbar{Peak energy rate powerlaw slope: $\alpha_P$}&&&		\\
Theory:  $\alpha_P$      &{\bf 1.00} &{\bf 1.50}	&{\bf 1.67}	\\
Simulation (N=256,64,24)&0.94 \pm 0.18	&1.85 \pm 0.06	&1.96 \pm 0.14	\\
Simulation (N=128,32,16)&1.05 \pm 0.10	&1.73 \pm 0.08	&1.95 \pm 0.13	\\
Charbonneau et al.~(2001)&		&1.72 \pm 0.02	&1.90 \pm 0.03	\\
\underbar{Duration powerlaw slope:  $\alpha_T$ }&&&		\\
Theory:  $\alpha_T$       &{\bf 1.00}&{\bf 1.5}     &{\bf 2.00}	\\
Simulation (N=256,64,24)&1.17 \pm 0.02	&1.77 \pm 0.18	&1.76 \pm 0.19	\\
Simulation (N=128,32,16)&1.27 \pm 0.15	&1.72 \pm 0.10	&1.76 \pm 0.18	\\
Charbonneau et al.~(2001)&		&1.71 \pm 0.01	&1.74 \pm 0.06	\\
\hline
\end{array}
$$
\end{table*}

\subsection{Occurrence Frequency Distributions}			 

The results of occurrence frequency distributions and correlations
are shown for a 1-D (Fig.~6), a 2-D (Fig.~7), and a 3-D cellular 
automaton code (Fig.~8), and listed in Table 2. 
We evaluated the powerlaw slopes by a weighted
linear regression fit (weighted by the number of avalanche events per bin)
by excluding undersampled bins (less than 20 events).
Since the exact values depend sometimes on the fitted range, we vary the
range of linear regression fits from the full range of the powerlaw part
(indicated with black line in Figs.~6-8) to the upper half
(indicated with grey line in Figs.~6-8) and calculate the average and 
standard deviations of the powerlaw slope values. 

The values of the predicted and simulated powerlaw slopes are compiled
in Table 2. First of all we notice an extremely good agreement (within
a few percents) of the powerlaw slopes $\alpha_E$, $\alpha_P$, and 
$\alpha_T$ between our simulations and those of Charbonneau et al.~(2001) 
and McIntosh et al.~(2002), although we used different codes, initiation 
times $t_{SOC}$, and powerlaw fitting procedures. We performed the 1-D
to 3-D runs with small cubes ($N=128, 32, 16$) as well as with larger
cubes ($N=256, 64, 24$), which both give very consistent values (Table 2).
In the following we quote only the values for the larger cubes, which
are also shown in Figs.~6-8.

Considering the agreement between theory and simulations,
there is a reasonable agreement for all three space dimensions $S=1,2,3$.
For the fractal dimension of avalanches we find $D_2 = 1.60 \pm0.17$
(predicted $D_2=1.5$) and $D_3=1.94\pm0.27$ (predicted $D_3=2.0$),
which are fully consistent with our theoretical estimate of
$D_S=(1+S)/2$ (Eq.~21). 

The relationship of the avalanche probability being reciprocal to the volume,
$N(L) \propto L^{-S}$ (Eq.~12) is also approximately confirmed by the 
simulations, for which
we find $\alpha_L=0.88\pm0.09$ (predicted $\alpha_L=1$ for $S=1$),
$\alpha_L=2.14\pm0.18$ (predicted $\alpha_L=2$ for $S=2$), and
$\alpha_L=2.55\pm0.10$ (predicted $\alpha_L=3$ for $S=3$), where the
latter value has the largest error due to the smallest sizes of 3-D cubes
with a dynamic range of only about 1 dex. Clearly this relationship
agrees better for larger cubes (up to $L=256$ in 1-D simulations).
We have to add a caveat that finite-size effects are likely to restrict
the size of avalanches at the system boundaries, especially for the 3D 
cellular automata runs which we run here with a size of $16^3$ and $24^3$ 
only. Also uncertainty estimates of the powerlaw slopes are less accurate
for small dynamic ranges (i.e., small system size $L$ here) and due to
histogram binning. In the real world, however, finite-size effects may
also cause modifications of powerlaw distributions, such as the maximum
size of active regions or the vertical density scale height of coronal
loops.   

The total energy powerlaw slope $\alpha_E$ (Eq.~18) fitted from the
powerlaw distributions $N(E)$ yields values of
$\alpha_E=1.06\pm0.04$ in 1-D (predicted $\alpha_E=1.0$),
$\alpha_E=1.48\pm0.03$ in 2-D (predicted $\alpha_E=1.28$), and
$\alpha_E=1.50\pm0.06$ in 3-D (predicted $\alpha_E=1.5$),
which agree with the predictions within 6\%, 16\%, and 0\%.

The peak energy dissipation rate powerlaw slope 
$\alpha_P$ (Eq.~18) fitted from the
powerlaw distributions $N(P)$ yields values of
$\alpha_P=0.94\pm0.18$ in 1-D (predicted $\alpha_P=1.0$),
$\alpha_P=1.85\pm0.06$ in 2-D (predicted $\alpha_P=1.5$), and
$\alpha_P=1.96\pm0.14$ in 3-D (predicted $\alpha_P=1.67$),
which agree with the predictions within 6\%, 23\%, and 17\%.

The time duration powerlaw slope $\alpha_T$ (Eq.~18) fitted from the
powerlaw distributions $N(T)$ yields values of
$\alpha_T=1.17\pm0.02$ in 1-D (predicted $\alpha_T=1.0$),
$\alpha_T=1.77\pm0.18$ in 2-D (predicted $\alpha_T=1.5$), and
$\alpha_T=1.76\pm0.19$ in 3-D (predicted $\alpha_T=2.0$),
which agree with the predictions within 17\%, 18\%, and 12\%.

Most occurrence frequency distributions are subject to a
drop-off from an ideal powerlaw distribution due to
finite-size effects. It is quite satisfactory that our simple 
first-order theory predicts most powerlaw slopes within an accuracy
of order 10\%. We have also to be aware that our first-order theory 
assumes that the fractal dimension of the energy release rate is a constant, 
while the simulated avalanches show a slight trend of increasing fractal 
dimensions $D$ with size $L$, i.e., $D_2 \approx L^{0.09}$ for $S=2$, 
and $D_3 \approx L^{0.19}$ for $S=3$ (Figs.~7 and 8, bottom right).
This slight trend of $D_S(L)$ affects the powerlaw slopes $\alpha_P$ 
and $\alpha_E$ to be somewhat flatter for small avalanches than for 
larger ones, which represents a second-order effect and could be 
considered in a more refined theory.

\begin{table*}
\caption{Theoretically predicted and numerically simulated 
correlations between the length scale $L$, time duration $T$, peak
energy dissipation rate $P$, and total energy $E$ for cellular 
automaton models with Euclidean dimension $S=1,2,3$ in the state
of self-organized criticality.}
$$
\begin{array}{p{0.5\linewidth}llll}
\hline\noalign{\smallskip}
Reference 	& S=1 	& S=2	& S=3	\\
\hline\noalign{\smallskip}
\hline
\hline\noalign{\smallskip}
\underbar{Fractal Dimension: $D_S$}&	&		&		\\
\underbar{Diffusive scaling of time with length: $L \propto T^{\beta_{TL}} = T^{1/2}$} &&&\\
Theory: $\beta_{TL}$    & {\bf 0.50}	&{\bf 0.50}	&{\bf 0.50}	\\
Slopes     (N=256,64,24)&        	&0.68 \pm 0.25	&0.49 \pm 0.21	\\
Slopes     (N=128,32,16)&		&0.76 \pm 0.13	&0.48 \pm 0.21	\\
Regression (N=256,64,24)&0.53 \pm 0.15	&0.65 \pm 0.02	&0.53 \pm 0.02	\\
Regression (N=128,32,16)&0.65 \pm 0.13	&0.61 \pm 0.01	&0.53 \pm 0.02	\\
\underbar{Correlation of power with duration: $P \propto T^{\beta_{TP}}$} &&&\\
Theory: $\beta_{TP}$    & {\bf 0.50}	&{\bf 1.00}	&{\bf 1.50}	\\
Slopes     (N=256,64,24)&		&0.91 \pm 0.19	&0.79 \pm 0.24	\\
Slopes     (N=128,32,16)&		&0.99 \pm 0.12	&0.80 \pm 0.22	\\
Regression (N=256,64,24)&0.67 \pm 0.20	&1.04 \pm 0.11	&1.01 \pm 0.12	\\
Regression (N=128,32,16)&0.73 \pm 0.14	&1.01 \pm 0.08	&1.02 \pm 0.14	\\
\underbar{Correlation of energy with duration: $E \propto T^{\beta_{TE}}$} &&&\\
Theory: $\beta_{TE}$    & {\bf 1.50} 	&{\bf 1.75}	&{\bf 2.00}	\\
Slopes     (N=256,64,24)&		&1.61 \pm 0.08	&1.51 \pm 0.20	\\
Slopes     (N=128,32,16)&		&1.73 \pm 0.10	&1.48 \pm 0.19	\\
Regression (N=256,64,24)&1.56 \pm 0.09	&1.80 \pm 0.05	&1.75 \pm 0.06	\\
Regression (N=128,32,16)&1.65 \pm 0.07	&1.82 \pm 0.04	&1.75 \pm 0.06	\\
\hline
\end{array}
$$
\end{table*}

\subsection{SOC Parameter Correlations}			 

An alternative method of testing our theory is (i) a determination of the
power index $\beta$ of correlated parameters $x$ and $y$,
i.e., $y \propto x^{\beta}$, by linear regression fits 
$\log(y) \propto \log(y_0)+\beta \log(x)$, or (ii) by inferring
them from the powerlaw slopes $\alpha_x$ and $\alpha_y$ of their 
occurrence frequency distributions, i.e., $\beta=(\alpha_x-1)/(\alpha_y-1)$
(see derivation in Section 7.1.6 of Aschwanden 2011). The resulting
values are listed for both methods in Table 3 (labeled with 
``Regression'' and ``Slopes''), for each 1-D, 2-D, and 3-D simulation
run of our cellular automaton code. The linear regression fits
are shown in the lower halves of the Figs.~6, 7, and 8. Note that
truncations occurs for each parameter due to the effect of finite-size 
systems that have been used in the numerical simulations ($L=$128 and 256
for 1-D; $L=$32 and 64 for 2-D; and $L=$16 and 24 for 3-D lattices).
The minimum amount of dissipated energy for a threshold of $B_c=5$ is 
$E_{min}=16.7$ (for $S=1$), $E_{min}=20$ (for $S=2$), and $E_{min}=21.4$ 
(for $S=3$), according to Eq.~(26). Thus we perform the linear
regression fits only in parameter ranges that are not too strongly
affected by truncation effects, which is at durations $T > 5$, 
peak energy dissipation rate $P > 50$, and energies $E > 50$. 
To quantify an error
of the linear regression fits, we perform fits of $y(x)$, and with
exchanged axes, $x(y)$, and quote the mean and half difference for
the two fits.

The correlation $L \propto T^{1/2}$ (Eq.~4) tests our assumption of a diffusive 
random walk for the propagating avalanche boundaries. For the theoretically
expected value of the power index $\beta_{TL}=0.5$ we find
$\beta_{TL}=0.53-0.65$ for 1-D avalanches,
$\beta_{TL}=0.61-0.76$ for 2-D avalanches, and 
$\beta_{TL}=0.48-0.53$ for 3-D avalanches,  
which corroborates our assumption of a diffusive avalanche expansion.

For the correlation of the peak energy dissipation rate
with the duration
of an avalanche $P \propto T^{\beta_{TP}}$ we find
$\beta_{TP}=0.67-0.73$ for 1-D avalanches (predicted $\beta_{TP}=0.5$),
$\beta_{TP}=0.91-1.04$ for 2-D avalanches (predicted $\beta_{TP}=1.0$), and
$\beta_{TP}=0.79-1.02$ for 3-D avalanches (predicted $\beta_{TP}=1.5$),
which amounts to an agreement of $\approx 30\%$ for the 1-D case, 
$\approx 3\%$ for the 2-D case, and $\approx 33\%$ for the 3-D case.
We suspect that a lot of avalanches are stopped at the boundary,
which underestimates the peak energy dissipation rate
and thus yields a systematically too high power index $\beta_{TP}$ for 
the 3-D case.

For the correlation of the total energy with the duration
of an avalanche $E \propto T^{\beta_{TE}}$ we find
$\beta_{TE}=1.56-1.65$ for 1-D avalanches (predicted $\beta_{TP}=1.5$),
$\beta_{TE}=1.61-1.82$ for 2-D avalanches (predicted $\beta_{TP}=1.75$), and
$\beta_{TE}=1.51-1.75$ for 3-D avalanches (predicted $\beta_{TP}=2.0$),
which amounts to an agreement of $\approx 7\%$ for the 1-D case, 
$\approx 2\%$ for the 2-D case, and $\approx 18\%$ for the 3-D case.

In summary, we find an overall agreement of order 10\% between theory and
numerical simulations for the power indexes of correlated parameters.

\begin{figure*}
\centerline{\includegraphics[width=\textwidth]{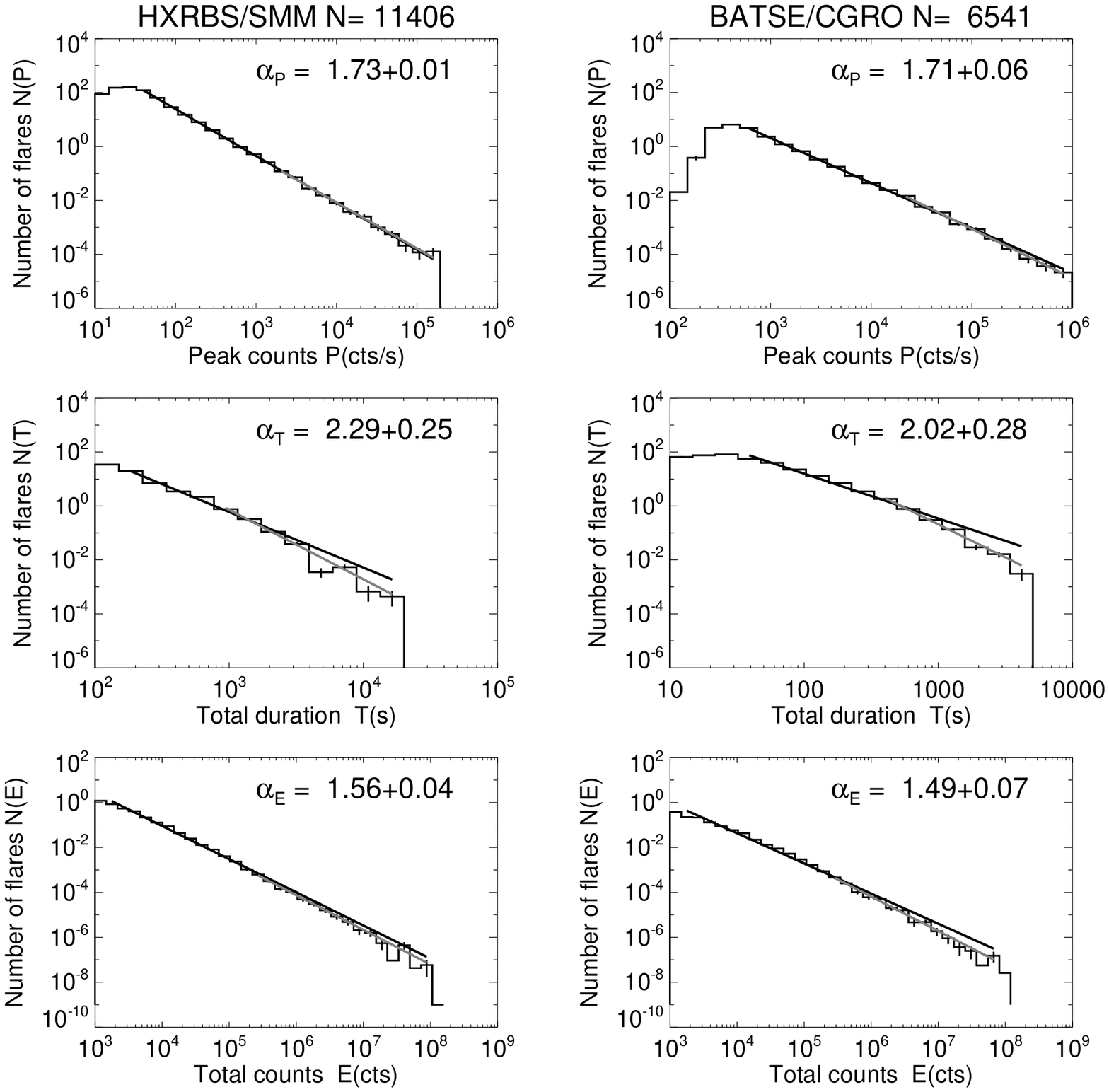}}
\caption{Occurrence frequency distributions of peak count rates $P$,
durations $T$, and total counts $E$ in solar flares observed with 
HXRBS/SMM during 1980-1989 (left panels ) and with BATSE/CGRO during 
1991-2000 (right panels) at energies $\ge25$ keV (left).}
\end{figure*}

\section{OBSERVATIONS OF SOLAR FLARES}

Let us turn now to some astrophysical observations to compare our theory
and the results of cellular automaton simulations. Lu and Hamilton (1991)
applied SOC theory and cellular automaton models for the first time to solar
flares. They modeled the flare statistics from hard X-ray counts observed
with the {\sl HXRBS} detectors onboard {\sl SMM}, which shows a powerlaw 
distribution extending over 4 orders of magnitude, with a powerlaw slope 
of $\alpha_P \approx 1.8$ for the peak count rate $P$ (Dennis 1985). 
From their cellular automaton code they obtained powerlaw slopes of 
$\alpha_P \approx 1.8$ for the peak energy dissipation rate $P$, 
$\alpha_T \approx 1.9$ for the flare durations $T$, and
$\alpha_E \approx 1.4$ for the total energies $E$. 
We re-analyzed the HXRBS/SMM dataset and obtain the values of 
$\alpha_P \approx 1.73\pm0.01$ for the power $P$, 
$\alpha_T \approx 2.29\pm0.25$ for the flare durations $T$, and
$\alpha_E \approx 1.56\pm0.04$ for the total energies $E$ (Fig.~9, left).
In addition we re-analyzed the BATSE/CGRO flare data set and obtained
similar values of
$\alpha_P \approx 1.71\pm0.06$ for the peak energy dissipation rate $P$, 
$\alpha_T \approx 2.02\pm0.28$ for the flare durations $T$, and
$\alpha_E \approx 1.49\pm0.07$ for the total energies $E$ (Fig.~9, right).

Comparing the observed values with the theoretically predicted
values of our 3-D {\sl fractal-diffusive SOC model} (Section 2 and Table 1),
we find the following ranges regarding the powerlaw slopes of 
the occurrence frequency distributions:
flare durations   $\alpha_T=2.02-2.29$ (predicted $\alpha_T^{theo}=2.00$), 
peak energy dissipate rate $\alpha_P=1.71-1.73$ (predicted $\alpha_P^{theo}=1.67$), and
total energies    $\alpha_E=1.49-1.56$ (predicted $\alpha_E^{theo}=1.50$), 
Thus, theory and observations agree within a few percents, which is even
better than the agreement of theory with the cellular automaton simulations
in the 3D case (probably due to the limited grid size).
The powerlaw index of the peak fluxes was found to
vary with the solar cycle within a range of $\alpha_P \approx 1.6-1.9$
(Crosby et al.~1993; Biesecker 1994; Bai 1993; Aschwanden 2010a),
which could indicate a variable degree of magnetic complexity that
is manifested either in a time-dependent fractal dimension $D_S$,
or threshold $B_c$ of SOC events (Aschwanden 2010a,b).

Comparing the observed flux time profiles $f(t)$ of the hard X-ray flux 
of large solar flares, they fluctuate erratically in a similar way as
shown for the largest avalanche of cellular automaton simulations (Fig.~3,
top left panel), as records with high time-resolution and high sensitivity
show, e.g., observed with BATSE/CGRO (Aschwanden et al.~1998). The
erratically fluctuating hard X-ray time profiles are generally interpreted
in terms of the chromospheric energy dissipation rate of precipitating
nonthermal electrons produced in magnetic reconnection processes, similar
as we defined the instantaneous energy dissipation rate $f(t)$ (Eq.~8) in 
our cellular automaton model, while the associated soft X-ray time profile 
represents the thermal emission of the heated plasma, which is monotonically
increasing during the impulsive flare phase because it represents the 
time integral of the heating rate according to the Neupert effect, similar 
as we defined the total dissipated energy $e(t)$ (Eq.~9) in our cellular
automaton model. The positive index in the correlation of the peak 
energy dissipation rate $P$ with the time duration $T$ in our avalanche
model, i.e., $P \propto T^{S/2}$ (Eq.~10), thus predicts that the hard
X-ray peak flux is statistically the higher the longer a solar flare lasts
(which is also known as ``big-flare syndrome'').  

A specific prediction is that in 3-D Euclidean space, regardless
of the fractal dimension in the range of $D_3=1, ..., 3$, the distribution of
flare energies is restricted to a relatively small range of 
$\alpha_E^{theo}=1.40-1.67$, which is consistent with our observations
of $\alpha_E=1.49-1.56$. The numerical value of
the energy powerlaw slope of $\alpha_E \approx 1.5$ implies that the
total energy of all flares is heavily weighted by the largest flares,
while nanoflares contain only an insignificant amount of energy,
unless the powerlaw slope is steeper than a critical value of 
$\alpha_E=2$ (Hudson 1991), which is an ongoing argument in the
controversy of coronal heating by nanoflares. In order to make
nanoflares dominant, a powerlaw slope of $\alpha_E > 2$ is needed,
which contradicts our model, since the maximum value cannot exceed
$\alpha_{E,max}=5/3=1.67$, well below the critical limit of $\alpha_E=2$.

{\bf Interestingly, some cellular automaton simulations have been performed
that actually produced significantly steeper powerlaw slopes,
say in the order of $\alpha_E \approx 3.0$, which 
seems to contradict our conclusions. One simulation produced a broken
powerlaw distribution with a slope of $\alpha_P=3.5$ for the smallest 
events, interpreted as nanoflare regime, while a flatter slope of
$\alpha_P=1.8$ was found for the larger flares (Vlahos et al.~1995),
a difference that is attributed to the anisotropic next-neighbor 
interactions applied therein. Another study simulated solar flare
events as cascades of reconnecting magnetic loops, with the finding
of a powerlaw slope of $\alpha_E=3.0$ for the released energies
(Hughes et al.~2003). Interacting loops with a length scale $L$ are 
bisected during a reconnection step into two shorter scales $L/2$, 
and the released energies are defined in terms of the length scale 
therein, i.e., $E \propto L$, for which our model predicts indeed
an occurrence frequency distribution of $N(E) \propto N(L) \propto
L^{-S} \propto L^{-3}$ (Eq.~12) for an Euclidean dimension $S=3$,
so it is fully consistent with our model if their energy definition
is adopted. Discrepancies of powerlaw slopes among different studies
can indeed often be explained in terms of inconsistent definitions
of the energy quantity.}

\section{CONCLUSIONS}

We developed an analytical theory for the statistical distributions
and correlations of observable parameters of SOC events, which includes
the avalanche length scale $L$, the time duration $T$, the 
peak $P$ and energy dissipation rate $F$.
The basic assumptions of our analytical model, 
which we call the {\sl fractal-diffusive SOC model}, are the following:

\begin{enumerate}
\item{{\bf Diffusive expansion of SOC avalanches:} The radius $r(t)$ 
	or spatial length scale $L$ of an avalanche grows with time
	like the average of a diffusive random walk, which predicts
	a statistical correlation $L \propto T^{1/2}$ between the
	length scale $L$ and time duration $T$ of the avalanche.}

\item{{\bf Fractal Energy Dissipation Rate:} The complexity of
	random next-neighbor interactions in a critical SOC state 
	can be characterized approximately with a fractal geometry.
	The volume (or area) of the instantaneous energy dissipation rate 
	is assumed to have a fractal dimension $D_S$. The predicted 
	statistical correlations are:
	$F \propto T^{D_S/2}$,
	$P \propto T^{S/2}$, and 
	$E \propto T^{1+D_S/2}$.} 

\item{{\bf Mean Fractal Dimension:} 
	The mean fractal dimension $D_S$ for different
	Euclidean space dimensions $S=1,2,3$ can be estimated from the
	arithmetic mean of the minimum dimension for a propagating
	avalanche, $D_{S,min} \approx 1$, and the maximum (Euclidean)
	dimension $D_{S,max} = S$, which yields $D_S \approx (1+S)/2$.}

\item{{\bf Occurrence Frequency Distributions:} Equal probability of
	avalanches with size $L$ at various spatial locations in a 
	uniform, slowly-driven SOC system predicts a probability distribution
	of $N(L) \propto L^{-S}$. A direct consequence of this
	assumption, together with the other assumptions made above,
	yields a powerlaw function $N(x) \propto x^{-\alpha_x}$ for the 
	occurrence frequency distributions of all parameters, 
	which is the hallmark of a SOC system. The predicted
	powerlaw indices are: 
	$\alpha_T=(1+S)/2$,  
	$\alpha_F=1+(S-1)/D_S$, 
	$\alpha_P=2-1/S$, and
	$\alpha_E=1+(S-1)/(D_S+2)$.
        Specifically, for applications to 3-D phenomena, absolute values 
	are predicted for the powerlaw slopes $\alpha_L=3$ and
	$\alpha_T=2$, and $\alpha_P=1.67$, while a range of 
	$\alpha_E=1.4,...,1.67$ is expected for any fractal
	dimension in the range of $1 \le D_3 \le 3$.} 
\end{enumerate}

We have validated our theory by a detailed comparison with a set of 
SOC simulations using a specific form of a cellular automaton 
avalanche model (connectivity, stability threshold, redistribution
rule, etc), and found a good agreement between theory and numerical
simulations, in the order of $\approx 10\%$ for the powerlaw slopes 
($\alpha_L, \alpha_T, \alpha_E, \alpha_P$), the power indices of 
correlated parameters ($\beta_{TL}, \beta_{TP}, \beta_{TE}$), and the
fractal dimensions $D_S$, for all three Euclidean space dimensions $S=1,2,3$. 
Yet, at its most general level our theory is saying that
the self-similarity of energy release statistics in such models
is a direct reflection of the fractal nature of avalanches. The SOC flare
model recently proposed by Morales \& Charbonneau (2008, 2009) offers
an interesting test of this conclusion. Their model, defined on a set
of initially parallel magnetic flux strands, contained in a plane and
subjected to random sideways
deformation, with instability and readjustment occurring when the crossing
angle of two flux strands exceeds some threshold angle. This model is thus
strongly anisotropic, with pseudo-local stability and redistribution,
in the sense that these operators now act on nearest-neighbors nodes
located along each flux strand, rather than in the immediate spatial
vicinity of the unstable sites. This is very different from the
isotropic Lu et al.~(1993)-type SOC model used here for validation.
Yet, the results compiled in Table 2
of Morales \& Charbonneau (2008) for their highest resolution simulations
reveal that the theoretical occurrence frequency distributions obtained herein,
do hold within the stated uncertainties on the power-law fits.
Likewise, the power indices of the correlated parameters 
also agree with theory within the inferred uncertainties.
This provides additional empirical support to our conjecture that the 
fractal-diffusive SOC model does represent a robust characterization 
of avalanche energy release in SOC systems in general. 
However, it should be remembered that the inferred scaling laws
are only valid for a slowly-driven SOC system, while alternative
SOC systems with time-variable drivers or non-stationary input rates
exhibit modified occurrence frequency and waiting time distributions 
(Charbonneau et al.~2001; Norman et al.~2001), which was also found
in solar observations extending over multiple solar cycles 
(Crosby et al.~1993; Biesecker 1994; Bai 1993; Aschwanden 2010a).

What other predictions can be made from our analytical model?
For SOC processes in 3-D space, which is probably the most common
application in the real world, the mean fractal dimension is
predicted to be $D_3\approx 2.0$, which can be tested by measurements 
of fractal dimensions in observations. The 20 largest solar flares 
observed with TRACE have been analyzed in this respect and an area fractal
dimension of $D_2=1.89\pm0.05$ was found at the flare peaks,
which translates into a value of $D_3=2.10\pm0.14$ if we use an 
anisotropic flare arcade model (Aschwanden and Aschwanden 2008a). 
The distribution of flare energies is predicted to have a powerlaw
slope of $\alpha_E=1.50$, which closely matches the observed statistics
of solar flare hard X-ray emission ($\alpha_E \approx 1.49-1.56$).
Since this value is undisputably below the critical limit $\alpha=2$
of the energy integral, the total released energy is contained in
the largest flares and thus rules out any significant nanoflare heating 
of the solar corona. Another prediction, that we did not test here with
solar flare data, is the diffusive flare size scaling.
Straightforward tests could be carried out by gathering statistics of
the flare size evolution during individual flares (which are predicted
to scale as $x(t) \propto t^{1/2}$), as well as from the statistics
of a large sample of flares, which is predicted to show a correlation
$L \propto T^{1/2}$. The application of our fractal-diffusive SOC
model to solar flares implies that the subsequent triggering of
local magnetic reconnection events during a flare occurs as a diffusive
random walk. A similar finding of diffusive random walk was also found
in the turbulent flows of magnetic bright points in the lanes between
photospheric granular convection cells (Lawrence et al.~2001). 
The spatio-temporal scaling of the diffusive random walk predicts also
the size, duration, and energy of the largest flare, which is likely
to be constrained by the size $L_{AR} \propto T_{max}^{1/2}$ of the 
largest active region.

\acknowledgements  
The author thanks the Paul Charbonneau for helpful discussions and
contributions. This work is partially supported by NASA grant 
NAG5-13490 and NASA TRACE contract NAS5-38099. 
We acknowledge access to solar mission data and flare catalogs from the 
{\sl Solar Data Analysis Center(SDAC)} at the NASA Goddard Space Flight 
Center (GSFC).


\section*{REFERENCES}

\def\ref#1{\par\noindent\hangindent1cm {#1}}

\ref{Aschwanden, M.J., Dennis, B.R., and Benz, A.O. 1998,
	\apj 497, 972-993.}
\ref{Aschwanden, M.J. 2004,
	{\sl Physics of the Solar Corona - An Introduction}.
	Springer/Praxis, New York, ISBN 3-540-22321-5, 842p.}
\ref{Aschwanden, M.J. 2010a,
	{\sl The state of self-organized criticality of the Sun
	during the last three solar cycles. I. Observations},
	\solphys (online first), DOI	10.1007/s11207-011-9755-0.}
\ref{Aschwanden, M.J. 2010b,
	{\sl The state of self-organized criticality of the Sun
	during the last three solar cycles. II. Theoretical model},
	\solphys (in press).}
\ref{Aschwanden,M.J. 2011,
	{\sl Self-Organized Criticality in Astrophysics - The Statistics
	of Nonlinear Processes in the Universe}, Springer/Praxis, 
	New York, ISBN 978-3-642-15000-5, hard-cover, 892p.}
\ref{Aschwanden, M.J. and Aschwanden, P.D. 2008a, \apj 674, 530.}
\ref{Aschwanden, M.J. and Aschwanden, P.D. 2008b, \apj 674, 544.}
\ref{Aschwanden, M.J., Dennis, B.R., and Benz, A.O. 1998, \apj 497, 972.}
\ref{Bai, T. 1993, \apj 404, 805.}
\ref{Bak, P., Tang, C., and Wiesenfeld, K. 1987, Phys. Rev. Lett. 59/27, 381.}
\ref{Bak, P., Tang, C., and Wiesenfeld, K. 1988, Phys. Rev. A 38/1, 364.}
\ref{Bak, P. and Chen, K. 1989, J Physics D 38, 5.}
\ref{Bak, P. 1996, {\sl How nature works}, Copernicus, Springer-Verlag, 
	New York.}
\ref{Biesecker, D.A. 1994, PhD Thesis, University of New Hampshire.}
\ref{Charbonneau, P., S.W. McIntosh, W.W., Liu, H.-L., and Bogdan, T.J. 2001,
  	\solphys 203, 321.}
\ref{Crosby, N.B., Aschwanden, M.J., and Dennis, B.R. 1993, \solphys 143, 275.}
\ref{Dennis, B.R. 1985, \solphys 100, 645.}
\ref{Fermi, E. 1949, Phys. Rev. Lett. 75, 1169.}
\ref{Gutenberg, B. and Richer, C.F. 1954,
       {\sl Seismicity of the Earth and Associated Phenomena},
       Princeton University Press, Princeton, NJ, p.310 (2nd ed.).}
\ref{Hudson, H.S. 1991, \solphys 133, 357.}
\ref{Hughes, D., Paczuski, M., Dendy, R.O., Heleander, P., and McClements,
	K.G. 2003, {\sl Phys. Rev. Lett.} 90, 131101.}
\ref{Lawrence, J.K., Cadavid, A.C., Ruzmaikin, A., and Berger, T.E. 2001,
	Phys. Rev. Lett. 86, 5894.}
\ref{Litvinenko, Y.E. 1998a, \aap 339, L57.}
\ref{Liu, H.L., Charbonneau, P., Pouquet, A., Bogdan, T. and McIntosh, S.
	2002, Phys. Rev. 66, 056111.}
\ref{Lu, E.T. and Hamilton, R.J. 1991, \apj 380, L89.}
\ref{Lu, E.T., Hamilton, R.J., McTiernan, J.M., and Bromund, K.R. 1993,
       \apj 412, 841.}
\ref{Mandelbrot, B.B. 1977,
       {\sl Fractals: form, chance, and dimension}, Translation of
       {\sl Les objects fractals}, W.H. Freeman, San Francisco.}
\ref{Mandelbrot, B.B. 1983, {\sl The fractal geometry of nature},
       W.H. Freeman, San Francisco.}
\ref{Mandelbrot, B.B. 1985, Physica Scripta 32, 257.}
\ref{McIntosh, S.W., Charbonneau, P., Bogdan, T.J., Liu, H.-L., and 
	Norman, J.P. 2002, Phys. Rev. E 65, 046125.}
\ref{Morales, L., and Charbonneau, P. 2008, \apj 682, 654.}
\ref{Morales, L., and Charbonneau, P. 2009, \apj 698, 1893.}
\ref{Norman, J.P., Charbonneau, P., McIntosh, S.W., and Liu, H.L.
	2001, \apj 557, 891.}
\ref{Sornette, D. 2004,
       {\sl Critical phenomena in natural sciences: chaos, fractals,
       self-organization and disorder: concepts and tools},
       Springer, Heidelberg, 528 p.}
\ref{Turcotte, D.L. 1999,
       {\sl Self-organized criticality},
       Rep. Prog. Phys. 62, 1377.}
\ref{Vlahos, L., Georgoulis, M., Kluiving, R., and Paschos, P. 1995,
	\aap 299, 897.}
\ref{Willis, J.C. and Yule, G.U. 1922, \nat 109, 177.}

\end{document}